\documentclass[twocolumn]{aastex7}
\usepackage{amsmath}

\usepackage{paralist}

\newcommand{\listitemsep}{0pt} 

\begin{document}
\title{Highly Efficient Selection of High-Redshift Emission-Line Galaxies for future DESI-like surveys with Deep Multi-band Imaging}

\correspondingauthor{Yoquelbin~Salcedo Hernandez}

\author[0000-0001-9897-576X, gname='Yoquelbin ', sname='Salcedo Hernandez ']{Yoquelbin~Salcedo Hernandez}
\affiliation{Department of Physics \& Astronomy and Pittsburgh Particle Physics, Astrophysics, and Cosmology Center (PITT PACC), University of Pittsburgh, 3941 O'Hara Street, Pittsburgh, PA 15260, USA}
\affiliation{Pittsburgh Particle Physics, Astrophysics, and Cosmology Center (PITT PACC), University of Pittsburgh, Pittsburgh, PA 15260, USA}
\email{yos47@pitt.edu}

\author[0000-0001-8684-2222, gname='Jeffrey A.', sname='Newman']{Jeffrey~A.~Newman}
\affiliation{Department of Physics \& Astronomy and Pittsburgh Particle Physics, Astrophysics, and Cosmology Center (PITT PACC), University of Pittsburgh, 3941 O'Hara Street, Pittsburgh, PA 15260, USA}
\affiliation{Pittsburgh Particle Physics, Astrophysics, and Cosmology Center (PITT PACC), University of Pittsburgh, Pittsburgh, PA 15260, USA}
\email{janewman@pitt.edu}

\author[0000-0001-8085-5890, gname='Brett H.', sname='Andrews']{B.~H.~Andrews}
\affiliation{Department of Physics \& Astronomy and Pittsburgh Particle Physics, Astrophysics, and Cosmology Center (PITT PACC), University of Pittsburgh, 3941 O'Hara Street, Pittsburgh, PA 15260, USA}
\affiliation{Pittsburgh Particle Physics, Astrophysics, and Cosmology Center (PITT PACC), University of Pittsburgh, Pittsburgh, PA 15260, USA}
\email{andrewsb@pitt.edu}

\author[0000-0002-5665-7912, gname='Biprateep', sname='Dey']{B.~Dey}
\affiliation{Department of Physics \& Astronomy and Pittsburgh Particle Physics, Astrophysics, and Cosmology Center (PITT PACC), University of Pittsburgh, 3941 O'Hara Street, Pittsburgh, PA 15260, USA}
\affiliation{Pittsburgh Particle Physics, Astrophysics, and Cosmology Center (PITT PACC), University of Pittsburgh, Pittsburgh, PA 15260, USA}
\affiliation{Department of Astronomy \& Astrophysics, University of Toronto, Toronto, ON M5S 3H4, Canada}
\email{b.dey@utoronto.ca}

\author[0000-0001-5381-4372, gname='Rongpu', sname='Zhou']{R.~Zhou}
\affiliation{Lawrence Berkeley National Laboratory, 1 Cyclotron Road, Berkeley, CA 94720, USA}
\email{rongpuzhou@lbl.gov}

\author[0000-0002-5333-8983, gname='Noah', sname='Sailer']{N.~Sailer}
\affiliation{Kavli Institute for Particle Astrophysics and Cosmology, 382 Via Pueblo Mall Stanford, CA 94305, USA}
\affiliation{SLAC National Accelerator Laboratory, 2575 Sand Hill Road, Menlo Park, CA 94025, USA}
\affiliation{University of California, Berkeley, 110 Sproul Hall \#5800 Berkeley, CA 94720, USA}
\affiliation{Lawrence Berkeley National Laboratory, 1 Cyclotron Road, Berkeley, CA 94720, USA}
\email{nsailer@stanford.edu}

\author[,gname='Jessica Nicole', sname='Aguilar']{J.~Aguilar}
\affiliation{Lawrence Berkeley National Laboratory, 1 Cyclotron Road, Berkeley, CA 94720, USA}
\email{jaguilar@lbl.gov}

\author[0000-0001-6098-7247, gname='Steven', sname='Ahlen']{S.~Ahlen}
\affiliation{Department of Physics, Boston University, 590 Commonwealth Avenue, Boston, MA 02215 USA}
\email{ahlen@bu.edu}

\author[0000-0001-9712-0006, gname='Davide', sname='Bianchi']{D.~Bianchi}
\affiliation{Dipartimento di Fisica ``Aldo Pontremoli'', Universit\`a degli Studi di Milano, Via Celoria 16, I-20133 Milano, Italy}
\affiliation{INAF-Osservatorio Astronomico di Brera, Via Brera 28, 20122 Milano, Italy}
\email{davide.bianchi1@unimi.it}

\author[,gname='David', sname='Brooks']{D.~Brooks}
\affiliation{Department of Physics \& Astronomy, University College London, Gower Street, London, WC1E 6BT, UK}
\email{david.brooks@ucl.ac.uk}

\author[,gname='Rebecca', sname='Canning']{R.~Canning}
\affiliation{Institute of Cosmology and Gravitation, University of Portsmouth, Dennis Sciama Building, Portsmouth, PO1 3FX, UK}
\email{becky.canning@port.ac.uk}

\author[0000-0001-7316-4573, gname='Francisco Javier', sname='Castander']{F.~J.~Castander}
\affiliation{Institut d'Estudis Espacials de Catalunya (IEEC), c/ Esteve Terradas 1, Edifici RDIT, Campus PMT-UPC, 08860 Castelldefels, Spain}
\affiliation{Institute of Space Sciences, ICE-CSIC, Campus UAB, Carrer de Can Magrans s/n, 08913 Bellaterra, Barcelona, Spain}
\email{fjc@ice.csic.es}

\author[0000-0001-8996-4874, gname='Edmond', sname='Chaussidon']{E.~Chaussidon}
\affiliation{Lawrence Berkeley National Laboratory, 1 Cyclotron Road, Berkeley, CA 94720, USA}
\email{echaussidon@lbl.gov}

\author[,gname='Todd', sname='Claybaugh']{T.~Claybaugh}
\affiliation{Lawrence Berkeley National Laboratory, 1 Cyclotron Road, Berkeley, CA 94720, USA}
\email{tmclaybaugh@lbl.gov}

\author[0000-0002-2169-0595, gname='Andrei', sname='Cuceu']{A.~Cuceu}
\affiliation{Lawrence Berkeley National Laboratory, 1 Cyclotron Road, Berkeley, CA 94720, USA}
\email{acuceu@lbl.gov}

\author[0000-0002-1769-1640, gname='Axel ', sname='de la Macorra']{A.~de la Macorra}
\affiliation{Instituto de F\'{\i}sica, Universidad Nacional Aut\'{o}noma de M\'{e}xico,  Circuito de la Investigaci\'{o}n Cient\'{\i}fica, Ciudad Universitaria, Cd. de M\'{e}xico  C.~P.~04510,  M\'{e}xico}
\email{macorra@fisica.unam.mx}

\author[0000-0002-4928-4003, gname='Arjun', sname='Dey']{A.~Dey}
\affiliation{NSF NOIRLab, 950 N. Cherry Ave., Tucson, AZ 85719, USA}
\email{arjun.dey@noirlab.edu}

\author[,gname='Peter', sname='Doel']{P.~Doel}
\affiliation{Department of Physics \& Astronomy, University College London, Gower Street, London, WC1E 6BT, UK}
\email{apd@star.ucl.ac.uk}

\author[0000-0003-4992-7854, gname='Simone', sname='Ferraro']{S.~Ferraro}
\affiliation{Lawrence Berkeley National Laboratory, 1 Cyclotron Road, Berkeley, CA 94720, USA}
\affiliation{University of California, Berkeley, 110 Sproul Hall \#5800 Berkeley, CA 94720, USA}
\email{sferraro@lbl.gov}

\author[0000-0002-3033-7312, gname='Andreu', sname='Font-Ribera']{A.~Font-Ribera}
\affiliation{Instituci\'{o} Catalana de Recerca i Estudis Avan\c{c}ats, Passeig de Llu\'{\i}s Companys, 23, 08010 Barcelona, Spain}
\affiliation{Institut de F\'{i}sica d’Altes Energies (IFAE), The Barcelona Institute of Science and Technology, Edifici Cn, Campus UAB, 08193, Bellaterra (Barcelona), Spain}
\email{afont@ifae.es}

\author[0000-0002-2890-3725, gname='Jaime E.', sname='Forero-Romero']{J.~E.~Forero-Romero}
\affiliation{Departamento de F\'isica, Universidad de los Andes, Cra. 1 No. 18A-10, Edificio Ip, CP 111711, Bogot\'a, Colombia}
\affiliation{Observatorio Astron\'omico, Universidad de los Andes, Cra. 1 No. 18A-10, Edificio H, CP 111711 Bogot\'a, Colombia}
\email{je.forero@uniandes.edu.co}

\author[0000-0001-9632-0815, gname='Enrique', sname='Gaztañaga']{E.~Gaztañaga}
\affiliation{Institut d'Estudis Espacials de Catalunya (IEEC), c/ Esteve Terradas 1, Edifici RDIT, Campus PMT-UPC, 08860 Castelldefels, Spain}
\affiliation{Institute of Cosmology and Gravitation, University of Portsmouth, Dennis Sciama Building, Portsmouth, PO1 3FX, UK}
\affiliation{Institute of Space Sciences, ICE-CSIC, Campus UAB, Carrer de Can Magrans s/n, 08913 Bellaterra, Barcelona, Spain}
\email{gaztanaga@gmail.com}

\author[0000-0003-3142-233X, gname='Satya ', sname='Gontcho A Gontcho']{S.~Gontcho A Gontcho}
\affiliation{University of Virginia, Department of Astronomy, Charlottesville, VA 22904, USA}
\email{satya@virginia.edu}

\author[,gname='Gaston', sname='Gutierrez']{G.~Gutierrez}
\affiliation{Fermi National Accelerator Laboratory, PO Box 500, Batavia, IL 60510, USA}
\email{gaston@fnal.gov}

\author[0000-0002-9136-9609, gname='Hiram K.', sname='Herrera-Alcantar']{H.~K.~Herrera-Alcantar}
\affiliation{Institut d'Astrophysique de Paris. 98 bis boulevard Arago. 75014 Paris, France}
\affiliation{IRFU, CEA, Universit\'{e} Paris-Saclay, F-91191 Gif-sur-Yvette, France}
\email{herreraa@iap.fr}

\author[0000-0003-0201-5241, gname='Dick', sname='Joyce']{R.~Joyce}
\affiliation{NSF NOIRLab, 950 N. Cherry Ave., Tucson, AZ 85719, USA}
\email{richard.joyce@noirlab.edu}

\author[0000-0002-0000-2394, gname='Stephanie', sname='Juneau']{S.~Juneau}
\affiliation{NSF NOIRLab, 950 N. Cherry Ave., Tucson, AZ 85719, USA}
\email{stephanie.juneau@noirlab.edu}

\author[,gname='Robert', sname='Kehoe']{R.~Kehoe}
\affiliation{Department of Physics, Southern Methodist University, 3215 Daniel Avenue, Dallas, TX 75275, USA}
\email{kehoe@physics.smu.edu}

\author[0000-0002-8828-5463, gname='David', sname='Kirkby']{D.~Kirkby}
\affiliation{Department of Physics and Astronomy, University of California, Irvine, 92697, USA}
\email{dkirkby@uci.edu}

\author[0000-0003-3510-7134, gname='Theodore', sname='Kisner']{T.~Kisner}
\affiliation{Lawrence Berkeley National Laboratory, 1 Cyclotron Road, Berkeley, CA 94720, USA}
\email{tskisner@lbl.gov}

\author[0000-0001-6356-7424, gname='Anthony', sname='Kremin']{A.~Kremin}
\affiliation{Lawrence Berkeley National Laboratory, 1 Cyclotron Road, Berkeley, CA 94720, USA}
\email{akremin@lbl.gov}

\author[0000-0002-1134-9035, gname='Ofer', sname='Lahav']{O.~Lahav}
\affiliation{Department of Physics \& Astronomy, University College London, Gower Street, London, WC1E 6BT, UK}
\email{o.lahav@ucl.ac.uk}

\author[0000-0002-6731-9329, gname='Claire', sname='Lamman']{C.~Lamman}
\affiliation{The Ohio State University, Columbus, 43210 OH, USA}
\email{lamman.1@osu.edu}

\author[0000-0003-1838-8528, gname='Martin', sname='Landriau']{M.~Landriau}
\affiliation{Lawrence Berkeley National Laboratory, 1 Cyclotron Road, Berkeley, CA 94720, USA}
\email{mlandriau@lbl.gov}

\author[0000-0003-1887-1018, gname='Michael', sname='Levi']{M.~E.~Levi}
\affiliation{Lawrence Berkeley National Laboratory, 1 Cyclotron Road, Berkeley, CA 94720, USA}
\email{melevi@lbl.gov}

\author[0000-0003-4962-8934, gname='Marc', sname='Manera']{M.~Manera}
\affiliation{Departament de F\'{i}sica, Serra H\'{u}nter, Universitat Aut\`{o}noma de Barcelona, 08193 Bellaterra (Barcelona), Spain}
\affiliation{Institut de F\'{i}sica d’Altes Energies (IFAE), The Barcelona Institute of Science and Technology, Edifici Cn, Campus UAB, 08193, Bellaterra (Barcelona), Spain}
\email{mmanera@ifae.es}

\author[0000-0002-1125-7384, gname='Aaron', sname='Meisner']{A.~Meisner}
\affiliation{NSF NOIRLab, 950 N. Cherry Ave., Tucson, AZ 85719, USA}
\email{aaron.meisner@noirlab.edu}

\author[,gname='Ramon', sname='Miquel']{R.~Miquel}
\affiliation{Instituci\'{o} Catalana de Recerca i Estudis Avan\c{c}ats, Passeig de Llu\'{\i}s Companys, 23, 08010 Barcelona, Spain}
\affiliation{Institut de F\'{i}sica d’Altes Energies (IFAE), The Barcelona Institute of Science and Technology, Edifici Cn, Campus UAB, 08193, Bellaterra (Barcelona), Spain}
\email{rmiquel@ifae.es}

\author[0000-0002-2733-4559, gname='John', sname='Moustakas']{J.~Moustakas}
\affiliation{Department of Physics and Astronomy, Siena University, 515 Loudon Road, Loudonville, NY 12211, USA}
\email{jmoustakas@siena.edu}

\author[0000-0001-9070-3102, gname='Seshadri', sname='Nadathur']{S.~Nadathur}
\affiliation{Institute of Cosmology and Gravitation, University of Portsmouth, Dennis Sciama Building, Portsmouth, PO1 3FX, UK}
\email{seshadri.nadathur@port.ac.uk}

\author[0000-0003-3188-784X, gname='Nathalie', sname='Palanque-Delabrouille']{N.~Palanque-Delabrouille}
\affiliation{IRFU, CEA, Universit\'{e} Paris-Saclay, F-91191 Gif-sur-Yvette, France}
\affiliation{Lawrence Berkeley National Laboratory, 1 Cyclotron Road, Berkeley, CA 94720, USA}
\email{npalanque-delabrouille@lbl.gov}

\author[0000-0002-0644-5727, gname='Will', sname='Percival']{W.~J.~Percival}
\affiliation{Department of Physics and Astronomy, University of Waterloo, 200 University Ave W, Waterloo, ON N2L 3G1, Canada}
\affiliation{Perimeter Institute for Theoretical Physics, 31 Caroline St. North, Waterloo, ON N2L 2Y5, Canada}
\affiliation{Waterloo Centre for Astrophysics, University of Waterloo, 200 University Ave W, Waterloo, ON N2L 3G1, Canada}
\email{will.percival@uwaterloo.ca}

\author[0000-0001-7145-8674, gname='Francisco', sname='Prada']{F.~Prada}
\affiliation{Instituto de Astrof\'{i}sica de Andaluc\'{i}a (CSIC), Glorieta de la Astronom\'{i}a, s/n, E-18008 Granada, Spain}
\email{fprada@iaa.es}

\author[0000-0001-6979-0125, gname='Ignasi', sname='Pérez-Ràfols']{I.~P\'erez-R\`afols}
\affiliation{Departament de F\'isica, EEBE, Universitat Polit\`ecnica de Catalunya, c/Eduard Maristany 10, 08930 Barcelona, Spain}
\email{ignasi.perez.rafols@upc.edu}

\author[0000-0001-5999-7923, gname='Anand', sname='Raichoor']{A.~Raichoor}
\affiliation{Lawrence Berkeley National Laboratory, 1 Cyclotron Road, Berkeley, CA 94720, USA}
\email{araichoor@lbl.gov}

\author[,gname='Graziano', sname='Rossi']{G.~Rossi}
\affiliation{Department of Physics and Astronomy, Sejong University, 209 Neungdong-ro, Gwangjin-gu, Seoul 05006, Republic of Korea}
\email{graziano@sejong.ac.kr}

\author[0000-0002-9646-8198, gname='Eusebio', sname='Sanchez']{E.~Sanchez}
\affiliation{CIEMAT, Avenida Complutense 40, E-28040 Madrid, Spain}
\email{eusebio.sanchez@ciemat.es}

\author[,gname='David', sname='Schlegel']{D.~Schlegel}
\affiliation{Lawrence Berkeley National Laboratory, 1 Cyclotron Road, Berkeley, CA 94720, USA}
\email{djschlegel@lbl.gov}

\author[,gname='Michael', sname='Schubnell']{M.~Schubnell}
\affiliation{Department of Physics, University of Michigan, 450 Church Street, Ann Arbor, MI 48109, USA}
\affiliation{University of Michigan, 500 S. State Street, Ann Arbor, MI 48109, USA}
\email{schubnel@umich.edu}

\author[0000-0002-6588-3508, gname='Hee-Jong', sname='Seo']{H.~Seo}
\affiliation{Department of Physics \& Astronomy, Ohio University, 139 University Terrace, Athens, OH 45701, USA}
\email{seoh@ohio.edu}

\author[0000-0002-3461-0320, gname='Joseph Harry', sname='Silber']{J.~Silber}
\affiliation{Lawrence Berkeley National Laboratory, 1 Cyclotron Road, Berkeley, CA 94720, USA}
\email{jhsilber@lbl.gov}

\author[,gname='David', sname='Sprayberry']{D.~Sprayberry}
\affiliation{NSF NOIRLab, 950 N. Cherry Ave., Tucson, AZ 85719, USA}
\email{david.sprayberry@noirlab.edu}

\author[0000-0003-1704-0781, gname='Gregory', sname='Tarlé']{G.~Tarl\'{e}}
\affiliation{University of Michigan, 500 S. State Street, Ann Arbor, MI 48109, USA}
\email{gtarle@umich.edu}

\author[,gname='Benjamin Alan', sname='Weaver']{B.~A.~Weaver}
\affiliation{NSF NOIRLab, 950 N. Cherry Ave., Tucson, AZ 85719, USA}
\email{benjamin.weaver@noirlab.edu}

\author[0000-0001-5146-8533, gname='Christophe', sname='Yèche']{C.~Yèche}
\affiliation{IRFU, CEA, Universit\'{e} Paris-Saclay, F-91191 Gif-sur-Yvette, France}
\email{christophe.yeche@cea.fr}

\author[0000-0002-6684-3997, gname='Hu', sname='Zou']{H.~Zou}
\affiliation{National Astronomical Observatories, Chinese Academy of Sciences, A20 Datun Road, Chaoyang District, Beijing, 100101, P.~R.~China}
\email{zouhu@nao.cas.cn}

\begin{abstract}
Emission-line galaxies (ELGs) are an important tracer of baryon acoustic oscillations (BAO) and large-scale structure (LSS) at $z > 1$. In this work, we investigate the feasibility of using deep wide-area multi-band imaging (e.g., from the Rubin Observatory) to efficiently select high redshift ELGs. Using Hyper Supreme-Cam $grizy$ photometry and COSMOS2020 many-band photometric redshifts, we designed simple color cuts guided by a probabilistic random forest classifier to select galaxies at $z = 1.1$--$1.6$. We then empirically tested and refined these color cuts using two samples of galaxies with deep spectroscopy and broad color coverage obtained with the Dark Energy Spectroscopic Instrument (DESI). Compared to DESI ELGs at $z = 1.1$--$1.6$, we achieve a higher redshift measurement success rate (89\%\ versus\ 69\%), a much higher correct redshift range success rate (84\%\ versus\ 34\%), and a far higher net surface density yield (1372 $\mathrm{deg^{-2}}$ versus\ 660 $\mathrm{deg^{-2}}$). Combining our sample with current DESI ELGs would increase the net ELG number density by a factor of $\sim3$, moving it out of the shot-noise limited regime and reducing the uncertainties on the BAO scale parameter at $z = 1.1$--$1.6$ by a factor of $\sim 2$.

\end{abstract}

\section{Introduction}
\label{sec:intro}

When the universe was still very young and hot, before the first atoms could form, baryons and photons were coupled into a single fluid. The radiation pressure associated with this fluid caused sound waves to propagate outward from higher-density regions, commonly known as baryon acoustic oscillations (BAO). The distance these sound waves could travel before radiation and matter decoupled imprints a characteristic scale upon the clustering of matter, enabling a powerful probe of the expansion history of our universe and the nature of dark energy using the late-time clustering of galaxies on the sky (e.g., see Section 4 of \citealt{Weinberg_2013}). Accurately detecting and interpreting the BAO signal typically involves measuring many hundreds of thousands of spectroscopic redshifts and measuring statistics that describe their clustering. The cosmology community has therefore endeavored to design galaxy redshift surveys with sufficient number densities, volume, and reliability of redshift measurements. In this paper, we will investigate new selection methods that take advantage of upcoming imaging datasets and could be used for future spectroscopic surveys.

The first BAO detection by \citet{Eisenstein_2005} was done using data from the Sloan Digital Sky Survey (SDSS) (\citealt{York_2000}, \citealt{Eisenstein_2001}). That same year, the BAO was also detected using data from the final 2dF Galaxy Redshift Survey \citep{Cole_2df_2005}. Over the last two decades, there has been significant progress in assembling massive spectroscopic surveys for measuring the BAO feature, some of which include WiggleZ: \citet{Drinkwater_2010}, BOSS: \citet{Dawson_2013}, and eBOSS: \citet{Dawson_2016}. These surveys have been able to place strong constraints on the dark energy equation of state between redshifts $0.1 < z < 1.0$.

To get even tighter cosmological constraints using BAO, galaxy redshift surveys must reach higher redshifts and have increased sky area coverage to probe larger volumes and span a greater fraction of cosmic history. For redshifts up to $z \sim 1$, luminous red galaxies (LRGs) have represented an ideal target class as they can be readily identified via the 4000 $\AA$ break \citep{Prakash_2016} and their redshifts can be determined robustly via the strong absorption features in their spectra \citep{Zhou_2023}. The primary target class at the highest redshifts (extending to $z \sim 4$) has been quasars, which are rarer but can be selected by ultraviolet or infrared excesses in their continuum out to very high redshift \citep{Myers_2015, Chaussidon_2023}. At intermediate redshifts, the primary target class in current surveys is emission-line galaxies (ELGs), the selection of which will be the focus of this paper.

ELGs exhibit very strong emission lines in their spectrum as they are selected at an epoch where the star formation rate (SFR) of the universe was much higher than it is today (for a recent review see \citealt{F_rster_Schreiber_2020}). An extremely useful feature that is common in the spectra of these galaxies is the [O II] doublet at 3726, 3729 \AA . This doublet provides a reliable method of measuring spectroscopic redshifts (hereafter spec-z or $z_\mathrm{spec}$) due to its strength and small wavelength separation. Furthermore, ELGs are ubiquitous over $1 < z < 1.6$, making them the ideal tracers for measuring the BAO signal at these redshifts.

The current state of the art survey for BAO cosmology is the Dark Energy Spectroscopic Instrument (DESI) \citep{desi_collab_experiment_design, desi_collab_overview, desi_survey_ops}. Roughly one-third of the targets will be comprised of ELGs at redshifts $0.6 < z < 1.6$ \citep{Raichoor_2023}. DESI recently released BAO distance scale measurements from the first two years of data using galaxy, quasar, and Lyman-$\alpha$ forest tracers \citep{desi_collab_dr1, desi_bao_cosmology_2024, desi_collab_bao_rsd, desi_dr2_bao}. Combined with either cosmic microwave background (CMB) or type Ia supernova measurements, DESI results prefer a time-varying dark energy equation of state, a direct tension with the current $\Lambda$CDM paradigm that will need to be confirmed or excluded with future DESI data.

DESI ELG\_LOP (DESI ELGs) targets were selected using photometry in the optical $g$, $r$, and $z$ using data from the DESI Legacy Imaging Surveys \citep{Dey_2019}. For ELGs, the resulting selections were affected by both the limited number of optical bands available and the targets having magnitudes near the $5\sigma$ depth of that imaging, making the sample selection sensitive to systematics that could affect depth. Consequently, DESI ELGs exhibit significant density fluctuations across the DESI footprint and a relatively low efficiency of selecting objects beyond the redshift range covered by LRGs, providing only around 660 objects $\mathrm{deg^{-2}}$ with robust redshift classifications between $1.1 < z < 1.6$ \citep{Raichoor_2023}. In comparison, the DESI LRG selection \citep{Zhou_2023}, which targets objects significantly brighter than the limits of the Legacy Survey photometry, exhibits a much more uniform density of targets across the sky. 

The utility of a more efficient selection algorithm for high-z ELGs that can take advantage of new, deeper imaging datasets is highlighted in \citet{dawson2022}. In particular, a more efficient ELG selection would improve clustering statistics at high redshifts with new DESI spectroscopy. Better imaging is now available through wide-field surveys like Hyper Supreme-Cam Subaru Strategic Program (HSC; \citealt{Aihara_2017}) and will soon be available from the Vera C. Rubin Observatory Legacy Survey of Space and Time (LSST; \citealt{2019ApJ...873..111I}), providing more depth and additional photometric bands that are invaluable in correctly identifying spectral features to enable target selection of ELGs at high redshift. A cleaner and denser selection of DESI ELGs using HSC or LSST imaging would allow a DESI-II survey to provide definitive BAO cosmology constraints at redshifts $1.1 < z < 1.6$.

In this paper we use existing datasets with LSST-like depth to develop a more efficient ELG target selection algorithm and test the feasibility of increasing the current DESI ELG sample by roughly a factor of $\sim3$ via a future DESI-II survey \citep{DESI2}. We provide a detailed description of the BAO distance scale improvements that would be obtained from larger samples in Section \ref{sec:forecasts}. In Section \ref{sec:observations}, we discuss the photometric and spectroscopic data used. In Section \ref{sec:optimization}, we optimize and finalize the selection cuts for our spectroscopic sample and compare directly to results from DESI. Finally, we summarize our work and findings in Section \ref{sec:Summary / Future  Tests}.

\section{Potential Impacts Of Enlarging the DESI ELG sample}
\label{sec:forecasts}

In this section, we describe the improvements to BAO measurement uncertainties that would result from enlarging the DESI ELG sample at redshifts $1.1 < z < 1.6$.
We have estimated these improvements using the \verb|FishLSS|\footnote{\href{https://github.com/NoahSailer/FishLSS}{https://github.com/NoahSailer/FishLSS}} code, which has been shown to agree well with the traditional forecasting method of \cite{2007ApJ...665...14S}. More specifically, we take Table 2.3 of \citet{DESI:2016fyo} as the ELG number density that would result from the full DESI survey and approximate the ELG linear bias as $b(z) = 0.84/D(z)$, where $D(z)$ is the growth factor normalized to unity today.\footnote{We hold the linear bias fixed to its fiducial value when rescaling the ELG number density, but note that the linear bias should decrease modestly if lower-luminosity but still star-forming objects are included \citep{Coil_2008}.} We split the ELG sample into four linearly-spaced redshift bins with $\Delta z=0.2$ covering the range $0.7 < z < 1.5$ and assume a total 14,000 ${\rm deg}^2$ sky coverage. We note that using smaller area would result in less gain than what is reported in this work. In each redshift bin the errors on the BAO distance parameters ($\alpha_\perp$ and $\alpha_\parallel$) are derived from the post-reconstruction power spectrum with the ``Rec-Sym'' convention using the Zel'dovich approximation \citep{Chen:2019lpf}. Here, $\alpha_\perp$ and $\alpha_\parallel$ correspond to separations transverse to the line of sight that depend on differences in angle and separations along the line of sight that depend on differences in redshift respectively. We fix the cosmology to the fiducial values listed in Table 2 of \citet{Sailer:2021yzm} and marginalize over the linear bias parameter and the coefficients of 15 polynomials of the form $k^n \mu^{2m}$ (for $n=0,1,2,3,4$ and $m=0,1,2$) to remove broad-band information. The fiducial values of higher-order bias parameters and the scale cuts used when computing the Fisher matrix are all identical to those used in \citet{Sailer:2021yzm}. 

\begin{figure}[htp]
    \centering
    \includegraphics[width=\columnwidth]{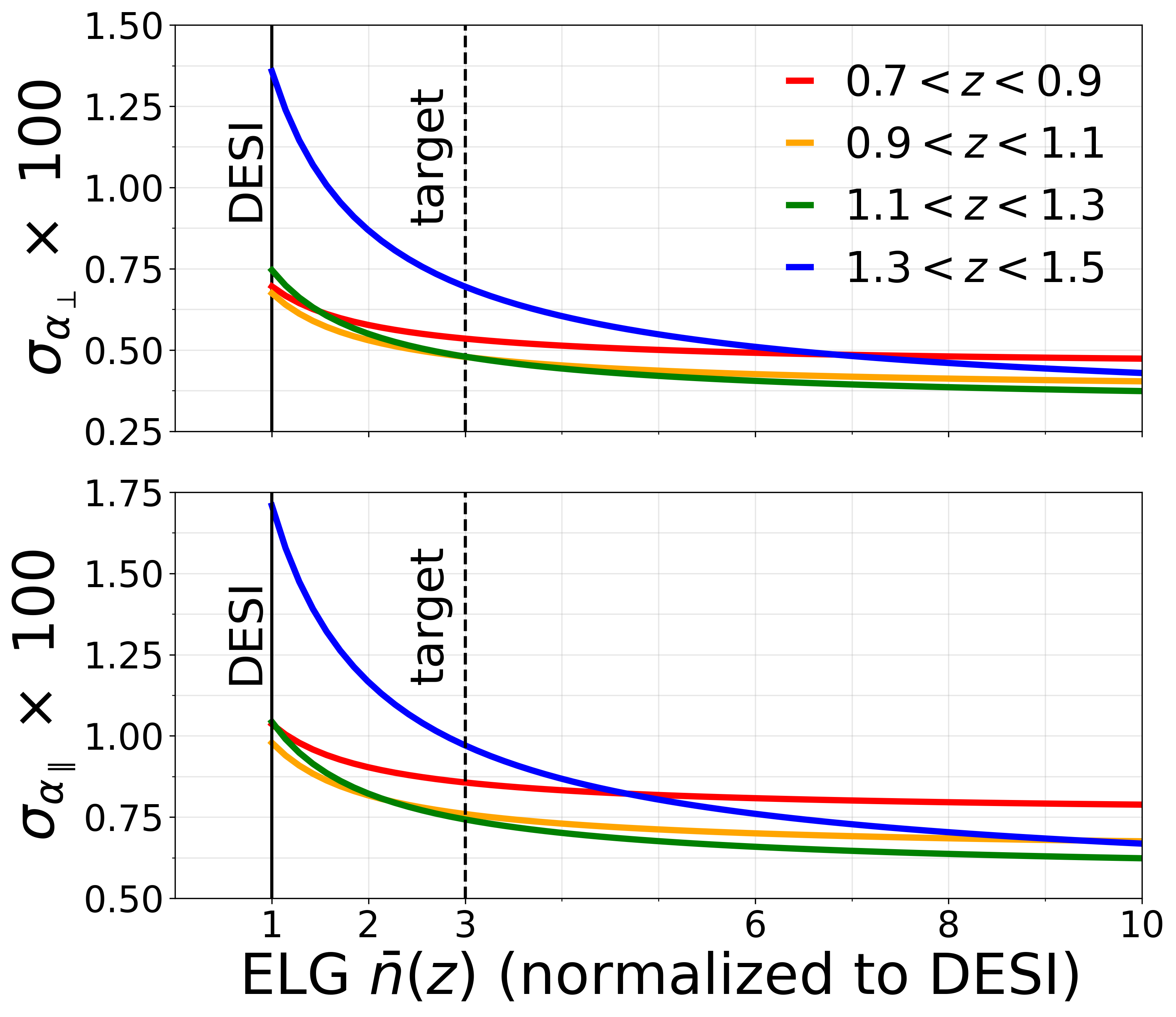}
    \caption{$1\sigma$ uncertainty on the BAO scale parameter $\alpha$ as a function of the factor by which the ELG number density is increased compared to the expected final DESI sample (solid black), for measures of both the transverse scale ($\alpha_{\perp}$) and the line of sight scale ($\alpha_{\parallel}$). The biggest gains are found for the $1.3 < z < 1.5$ redshift bin (blue curve), where expanding the ELG sample at these redshifts by $\sim3\times$ (indicated by the dashed black line) would yield a factor of $\sim2\times$ improvement on $\alpha_{\perp}$ and $\alpha_{\parallel}$. Enlarging the sample by more than this factor would yield only marginal additional improvements.
    }
    \label{fig:sailer}
\end{figure}

The resulting estimates of the improvements on the uncertainties of $\alpha_{\perp}$ and $\alpha_{\parallel}$ as a function of the factor by which the ELG number density increases compared to the full DESI survey are shown in Figure \ref{fig:sailer}. The solid black vertical line denotes the expectation for the final DESI ELG sample for comparison. For the three lowest-redshift bins gains are quite modest for expansions of the sample by  $>2\times$. In contrast, for the highest redshift bin ($1.3 < z < 1.5$, shown in blue), gains only become slow when the sample is expanded by $>3\times$, a limit marked by the black dashed vertical line. Increasing the number density by more than this would result in only marginal decreases in the error on either parameter. Given this, we want to obtain a sample with a surface density yield of 1370 $\mathrm{deg^{-2}}$, compared to 660 $\mathrm{deg^{-2}}$ for the current DESI sample. These calculations were made based on the original 5-year DESI survey design. The change in area from 14000 $\mathrm{deg^{2}}$ to 17000 $\mathrm{deg^{2}}$ expected from the extended DESI program would lower each curve by a constant factor. It is also important to note that these gains are after fiber assignment, not before.

\section{Data} 
\label{sec:observations}

Based on the estimates in Section \ref{sec:forecasts}, we aimed to augment the current DESI ELG sample at $z > 1.1$ by a factor of $\sim3\times$ by combining the current sample with the one we optimize in this work. In this section, we describe the photometric and spectroscopic data used to design new ELG selections that accomplish this goal. We use wide imaging data from the third data release of HSC containing LSST-like $grizy$ photometry. We used LSST-like deep $ugrizy$ photometry from the COSMOS2020 catalog and the many-band LePHARE photo-z's. We used the DESI spectroscopic truth sample for our final selection optimization, which was designed for studying the imaging systematics of the ELG sample.

\subsection{HSC}
\label{subsec:hsc_data}
In this work, we used the wide imaging from the third data release (DR3) of HSC \citep{Aihara_2022}. This catalog includes $grizy$ photometry with point spread function (PSF) depths of 26.5/26.5/26.2/25.2/24.4 respectively. HSC wide is slightly deeper than LSST Y2 in some bands, but we demonstrate that this does not qualitatively affect the metrics of interest for selecting high-z ELGs in Appendix \ref{appendix:shallower_selection}. The HSC bands are around 2 mag deeper than the depths of the Legacy Survey imaging that was used for DESI ELG targets and have the addition of $i/y$. The deeper and additional bands are important for greatly improving the efficiency of ELG target selections. We used the Schlegel, Finkbeiner, and Davis extinction map for MW extinction corrections \citep{Schlegel_Davis_Finkbeiner_1998}.

\subsection{COSMOS2020}
\label{subsec:cosmos2020_data}

As one avenue for testing the viability of selecting a large sample of ELGs at $1.1 < z < 1.6$ using LSST-like photometry, we used the Farmer version of the COSMOS2020 catalog \citep{cosmos2020}. This catalog includes deep Canada France Hawaii Telescope (CFHT) Megacam $u$ \citep{Sawicki_2019} and HSC $grizy$ \citep{Aihara_2019} photometry reaching depths of 27.7/28.1/27.8/27.6/27.2/26.5 in a 2 $^{\prime\prime}$ aperture for each band, respectively. The corresponding PSF depths for the HSC imaging are 28.3/28.1/28.0/27.5/26.6 respectively.

In total, this catalog contains photometry from more than 30 passbands, reduced with matched-model algorithms using the \texttt{Tractor} software \citep{2016ascl.soft04008L}.  It also includes photometric redshifts (photo-z's) that have been derived using all of the available passbands using both the LePHARE \citep{2011ascl.soft08009A} and EAZY \citep{2008ApJ...686.1503B} spectral energy distribution template fitting algorithms. While both algorithms typically provide similar redshift estimates, we adopt the LePHARE photo-z's for our analysis because EAZY results exhibited anomalies at $z > 1.5$. In particular, EAZY produced unrealistically high photo-z's out to z $\sim$ 12. Although the many-band LePHARE photo-z's are much more reliable than photo-z's based only on broad-band optical photometry, they still exhibit larger uncertainties than spectroscopic redshifts and are only available within small areas (making cosmic variance more of an issue); in Appendix \ref{appendix:photoz_tests} we present tests of these redshifts for ELG-like samples with DESI spec-z measurements.

\subsection{Spectroscopic Data}
\label{subsec:spec_data}

\begin{figure*}
    \centering
    \includegraphics[width=\textwidth]{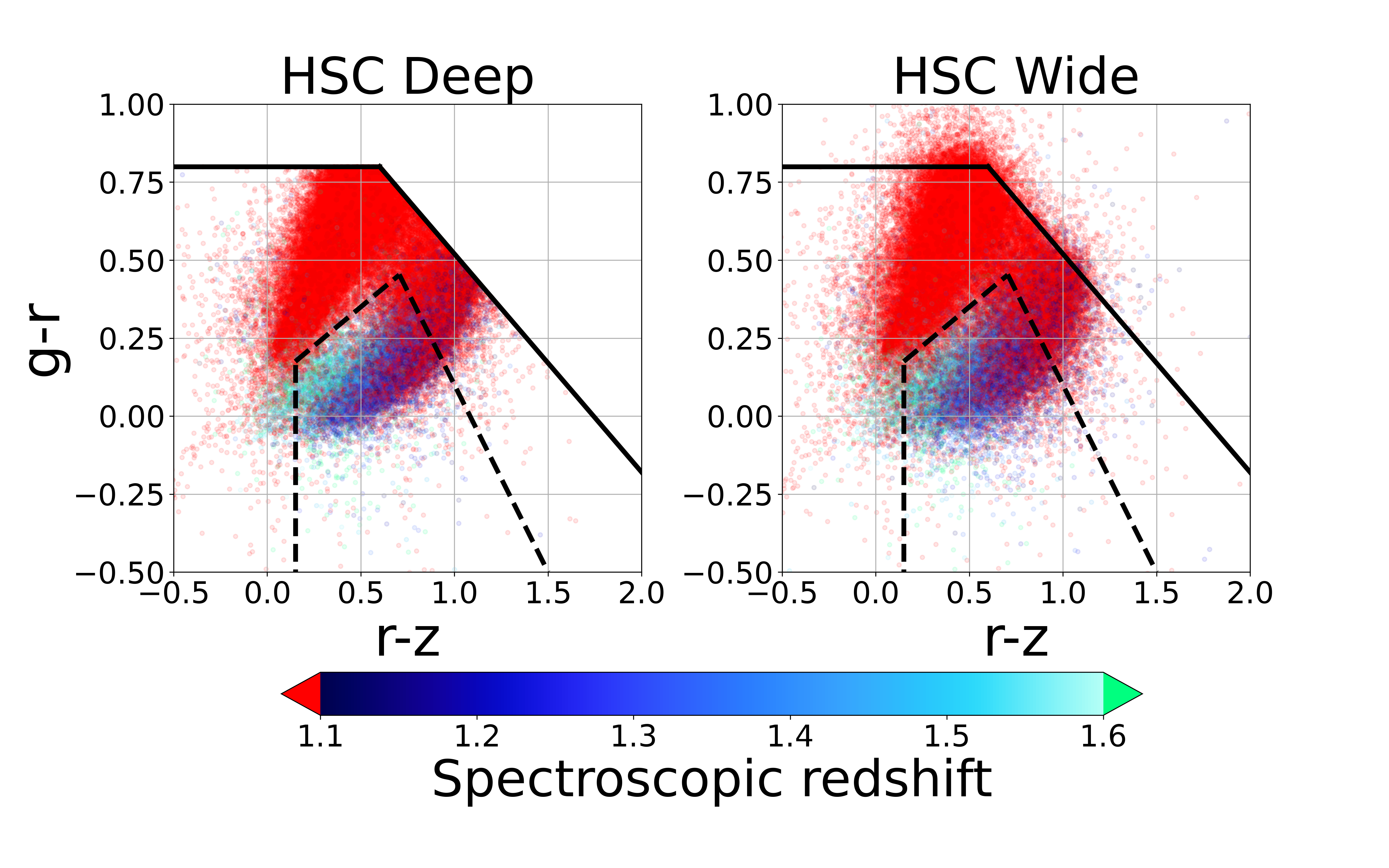}
    \caption{Color-color diagrams of the DESI spec-truth sample using $g-r/r-z$ plotted using imaging from either deep (left) or wide (right) depth HSC imaging, with galaxies color-coded according to their spec-z. This sample was obtained using HSC deep photometry. We show the color cuts for the standard DESI ELG sample (dashed black lines) and the spec-truth sample (solid black). Although the wide photometry is noisier, the selection limits are far enough from where ELGs reside in color space to ensure that the sample still retains all ELG-like objects down to the desired depths.
    \label{fig:targeted_color_color_truth}}
\end{figure*}

In addition to the COSMOS2020 sample, we also made use of a spectroscopic sample that was originally designed for studying the imaging systematics of the DESI ELG sample (Zhou et al; in prep.). This sample was targeted using deep DECam \citep{Flaugher_2015} and HSC imaging with selection cuts that extend well beyond those used to select DESI ELGs to capture all of the objects that could scatter into the DESI sample due to photometric uncertainties. From here on, we refer to this sample as the ``spec-truth'' sample. Figure \ref{fig:targeted_color_color_truth} shows the spec-truth (solid black lines) and DESI ELG (dashed black lines) color cuts in $g-r$ versus $r-z$ color space. The spec-truth sample was selected using HSC deep photometry from DR3 (left panel), but the broad color cuts of this sample should include all of the ELGs at $z = 1.1$--$1.6$ even when selecting objects using the shallower HSC wide photometry.

The selection cuts for the spec-truth sample using DECam photometry were
\begin{equation}
\tag{1a}
19.5 < g_\mathrm{fiber} < 24.5,  
\end{equation}

\begin{equation}
g - r < 0.9,
\tag{1b}
\end{equation} and

\begin{equation}
g - r < 1.32 - 0.7 \times (r-z),
\tag{1c}
\end{equation}

where $g_\mathrm{fiber}$ is the $g$-band fiber magnitude (i.e., the magnitude corresponding to the expected flux within a DESI fiber aperture). The spec-truth selection cuts applied based on HSC photometry were
\begin{equation}
\tag{2a}
19.5 < g_\mathrm{aperture} < 24.55, 
\end{equation}

\begin{equation}
g - r < 0.8,
\tag{2b}
\end{equation} and

\begin{equation}
g - r < 1.22 - 0.7\times (r-z),
\tag{2c}
\end{equation}

where $g_\mathrm{aperture}$ is the g-band magnitude within a 1.5 $^{\prime\prime}$ aperture.  Objects were included in the observed sample if they met either the HSC or DECam selection criteria. 

This sample was observed using the DESI instrument in the COSMOS field on March 9-13, 2024, utilizing 25 overlapping tiles on the sky with 1000 second effective exposure time for each tile \citep{desi_collab_experiment_design, desi_collab_overview, desi_collab_corrector, desi_collab_fiber_system}. We used the standard DESI definition of effective exposure time (see Equation 22 in \citealt{Guy_2023}). In total, spectra for 95,443 unique objects were obtained. After applying a minimum effective exposure time cut of $t > 700$ seconds and cross-matching with the HSC wide catalog, the total number of ELG-like objects available for our analysis was 87,512. 

\begin{figure}[htp]
    \centering
\includegraphics[width=\columnwidth]{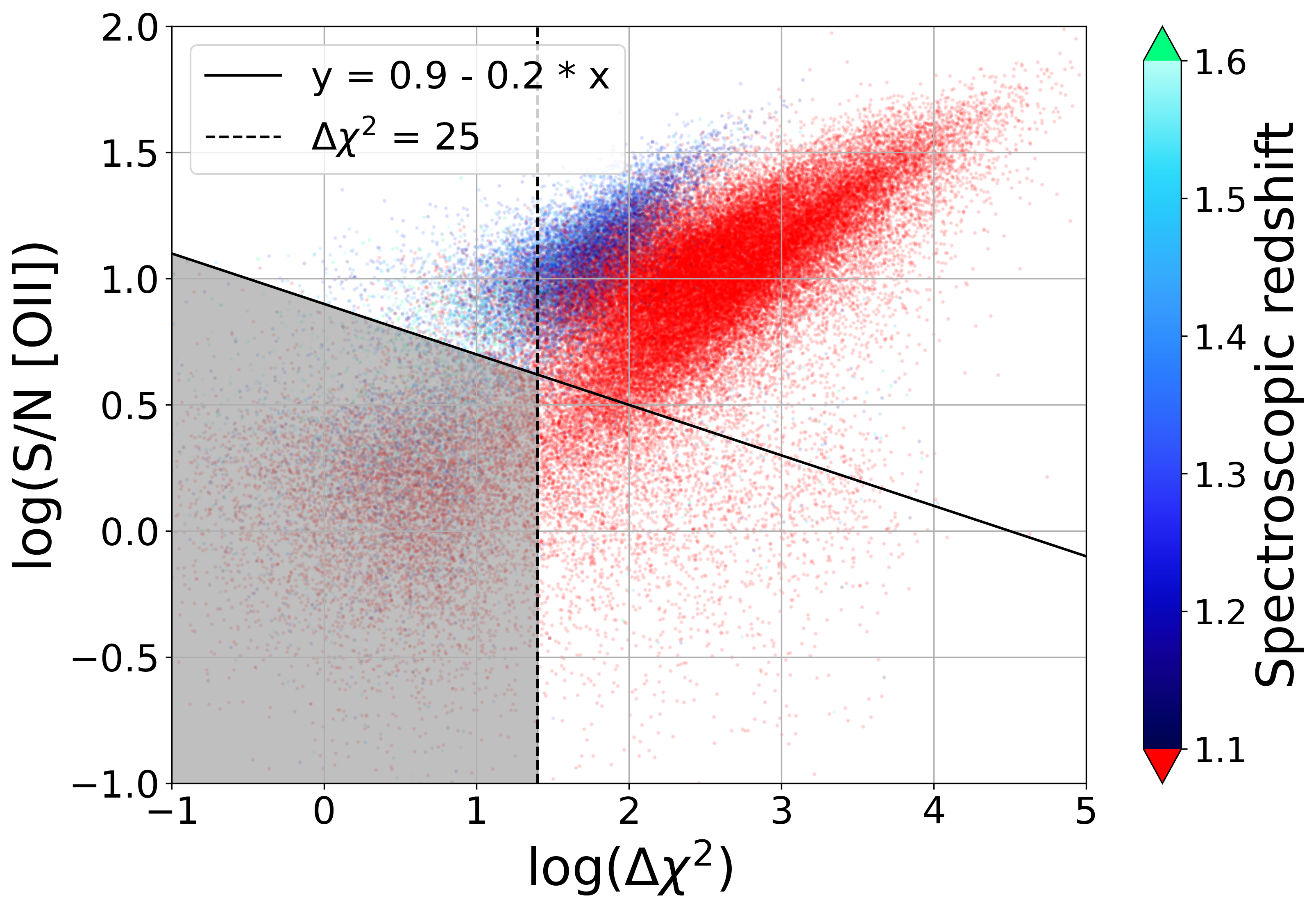}
\caption{
 [O II] flux S/N vs. the $\Delta\chi^2$ between the best-fit and second best-fit redshift solutions for the spec-truth sample (points color coded by spec-z). We consider objects above the solid black diagonal line (the DESI ELG reliable-redshift criteria; \citealt{Raichoor_2023}) or rightward of the dashed black line ($\Delta\chi^2$) to have reliable redshifts (i.e., not in the shaded region of this diagram).}
    \label{fig:good_z_criteria}
\end{figure}

In Figure \ref{fig:good_z_criteria}, we plot the [O II] flux signal-to-noise (S/N) vs. the $\Delta\chi^2$ between the best-fit and second best-fit redshift solutions for the spectroscopic truth sample (where points are colored by their spec-z). To select objects with reliable spec-z's, we applied similar cuts to those used for selecting DESI ELGs \citep{Raichoor_2023} based on the S/N of the flux of the [O II] $\lambda\lambda$ 3726, 3729 \AA \,emission line doublet and the $\Delta\chi^2$ between the best-fit and second best-fit spectral template. After determining that galaxies with $\Delta\chi^2 > 25$ below the solid black line should contain significant redshift information in their spectrum, we add this to our reliable-redshift cuts. The final reliable-redshift cuts used for this work are 

\begin{equation}
    \label{eq:reliable_z_criteria}
 \mathrm{log}(\text{flux}~[\mathrm{OII}]) > 0.9 - 0.2 \, \times \mathrm{log}(\Delta\chi^2) \, \textbf{OR} \,\Delta\chi^2 > 25, 
\end{equation}
         where the first term is the reliable-redshift cut used for DESI ELGs. We find that 68,035 of these galaxies have reliable spec-z measurements after applying the $\Delta\chi^2$ and [OII] S/N criteria cuts. These criteria constitute a more inclusive version of what was used in \citet{Raichoor_2023} for the final DESI ELG sample \footnote{If we instead only use the reliable-redshift criteria employed for DESI ELGs, we see a reduction of the redshift success rate from 89\% to 87\% and redshift range success rate from 84\% to 83\% for our final optimized selection.}.

\begin{figure}
    \centering
    \includegraphics[width=\columnwidth]{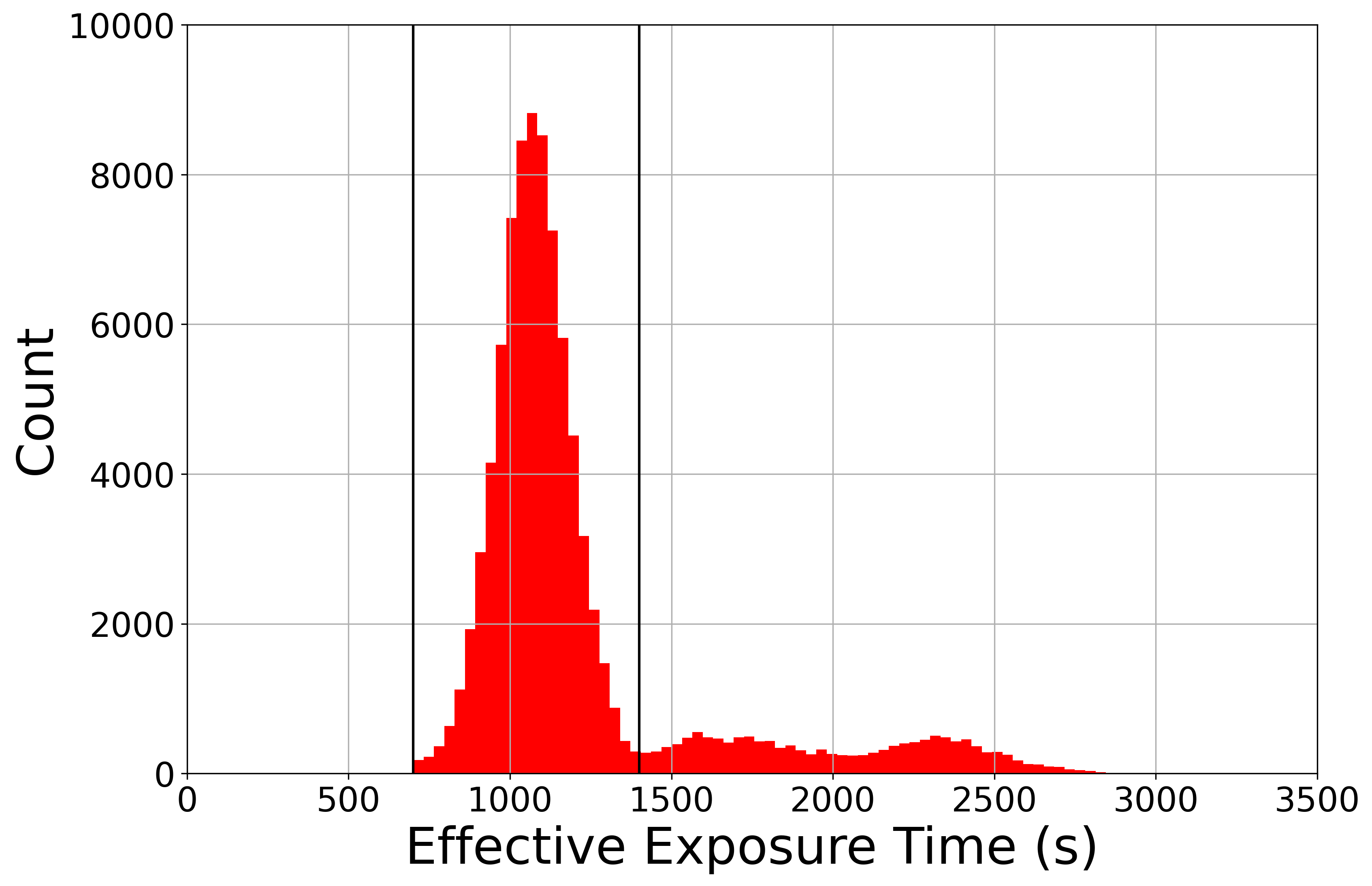}
    \caption{The effective exposure time distribution for the spec-truth sample. Only those objects whose exposure times were within the range bounded by the solid black vertical lines ( $700 < t < 1400$ seconds) were used for calculating redshift measurement success rates, whereas all objects with $t > 700$ seconds were included when calculating the fraction of targets in the desired redshift range ($1.1 < z < 1.6$). The metrics used are described in more detail in Section \ref{sec:optimization}.}
    \label{fig:exposure_truth}
\end{figure}

Figure \ref{fig:exposure_truth} shows the effective exposure time distribution for the spec-truth sample. In order to make as fair a comparison as possible to the DESI ELG sample, we only used objects with effective exposure times in the range $700 < t < 1400$ seconds (region between black solid vertical lines) for calculating redshift measurement success rate (i.e., how often a targeted object yielded a secure redshift measurement). However, we used all objects with effective exposure times above 700 seconds to calculate the fraction of objects at $1.1 < z < 1.6$.

\section{Design of Selection Criteria} 
\label{sec:optimization}
In this section we describe our method of acquiring initial color cuts and optimizing these cuts to cleanly select high redshift ELGs between $1.1 < z < 1.6$. In particular, we wanted to obtain a sample of ELGs with very high redshift measurement and redshift range success rates, in addition to our goal of a sample with a surface density yield of 1370 $\mathrm{deg^{-2}}$. We then compare our optimized sample to the current DESI ELGs over $1.1 < z < 1.6$.

\subsection{Initial Design of Color Cuts Using a Random Forest} \label{subsec:color_cut_design}
We used a random forest (RF; \citealt{randomforests}) to classify whether COSMOS2020 galaxies were high-z ELGs based on their $ugrizy$ photometry. RF classifiers deliver performance comparable to more sophisticated machine learning algorithms with minimal hyperparameter tuning and robustly handles degenerate information, making them well suited for our needs of developing a good initial set of color cuts. We trained the RF using the deep LSST-like COSMOS2020 $ugrizy$ magnitudes and all color combinations to predict which galaxies are within the range $z = 1.1$--$1.6$ based on the many-band LePHARE photo-z's.

The RF proved to be very successful at selecting high-z ELGs. Using a receiver operating characteristic (ROC; \citealt{2006PaReL..27..861F}) curve, a common tool for evaluating classifiers, the RF achieved an impressive area-under-the-ROC-curve of 0.960 (compared to 1 for a perfect classifier). This level of performance corresponds to a sample that is both highly pure and complete.

We then used the RF classifier probabilities (of being a high-z ELG) to design a set of color cuts to simplify the selection, ease simulation efforts, and maximize interpretability. First, we determined which colors were most influential on the RF classifier by determining the permutation feature importance for all input properties \citep{featureimportance}. We found that the combination of $r-i$ and $i-y$ colors were enough to capture most of the useful information. The next most important feature was $i-z$, which we considered for our final optimized sample (see Section \ref{subsec:colorcut_offset}). These colors are not unique as other color combinations exist that would contain degenerate information, which would lead to similarly effective selections. The color cuts were
\begin{itemize}
    \item $i-y - 0.19 \, +\mathrm{offset_{1}} > r-i$, 
    \item  $i-y > 0.35 \, +{\mathrm{offset}_{2}}$,
\end{itemize} where $\mathrm{offset_{1}}$ and $\mathrm{offset_{2}}$ are both zero (but are optimized in the next Subsection with the help of spectroscopy).

\subsection{Optimizing Color Cuts} \label{subsec:colorcut_offset}

After developing initial color cuts based upon COSMOS2020, we used the spec-truth sample to explore modifications to those cuts that would maximize the fraction of objects with secure redshifts in the desired redshift range  while yielding a net surface density yield close to our goal of 1370 $\rm{deg^{-2}}$, corresponding to our target of increasing the density of high-z ELGs by $\sim 3\times$ (see Section \ref{sec:forecasts}). In addition, we wanted the final sample to have a limiting magnitude not much fainter than the DESI ELG selection. We tested both $g$-fiber and $r$-fiber selections and found limiting with $g$-fiber to be better for selecting high-z ELGs. This is not too surprising as g-band is closer to the rest UV and so a better tracer of SFR and hence [O II] flux. To characterize and optimize the resulting samples, we considered several metrics: redshift measurement success rate, redshift range success rate, target density, and net surface density yield.

First, we define the redshift measurement success rate, \begin{equation} \label{eq:zrate}
f_\mathrm{reliable} = \frac{N_\mathrm{reliable}^{700\,<\, t\, <\,1400}}{N_\mathrm{total}^{700\,<\, t\, <\,1400}}, 
\end{equation} 
where ${N_\mathrm{total}^{700\,<\, t\, <\,1400}}$ is the number of objects with integration time between $t\, = \, 700\,-1400$ seconds that pass our optimized color cuts and $g$-fiber limiting magnitude, and $N_\mathrm{reliable}^{700\,<\, t\, <\,1400}$ are the subset of those objects that also pass the reliable $z_\mathrm{spec}$ cuts. The exposure time cuts were applied both $(1)$ to exclude objects with bad exposures or negligible effective observation time and $(2)$ to exclude objects with exposures significantly longer than a typical DESI ELG exposure \citep{Raichoor_2023}  to ensure that redshift measurement success rates could be compared fairly between samples.

To compute the redshift range success rate, we must first compute the fraction of all objects (with $ t > 700 $ seconds) with $z_\mathrm{spec} = 1.1$--$1.6$,  \begin{equation} 
\label{eq:zrangerate}
f_\mathrm{z\,= \, 1.1\, -\, 1.6}^{t\, >\, 700} = \frac{N_\mathrm{{z\,= \, 1.1\, -\, 1.6}}^{t\, >\, 700}}{N_\mathrm{total}^{t\, >\, 700}},
\end{equation} where ${N_\mathrm{total}^{t\, >\, 700}}$ is the total number of objects with an integration time of $t > 700$ seconds that pass our optimized color cuts and $g$-fiber limiting magnitude and have reliable $z_\mathrm{spec}$, and ${N_\mathrm{{z\,= \, 1.1\, -\, 1.6}}^{t\, >\, 700}}$ is the subset that are within $1.1 < z < 1.6$.  In this case, we include information from objects with exposure times longer than those typically used for DESI ELGs to maximize the sample size; since we are not assessing the redshift measurement success rate in this factor, a maximum exposure time cut is not needed. 

Now we can calculate the redshift range success rate; (i.e., the fraction of targets with typical DESI ELG exposure times that are in the desired redshift range with successfully measured $z_\mathrm{spec}$) \begin{equation} \label{eq:zrangeyield}
f_\mathrm{z\,= \, 1.1\, -\, 1.6}^{700 \,<\,t\, <\,1400} = f_\mathrm{reliable}\times f_\mathrm{z\,= \, 1.1\, -\, 1.6}^{t\, >\, 700},
\end{equation} where $f_\mathrm{reliable}$ accounts for redshift measurement success rate for $700 < t < 1400$ second exposures only.

We calculate the target density, \begin{equation} \label{eq:target_density}
\Sigma_\mathrm{target} = \frac{N_\mathrm{selected}}{A},
\end{equation} 
where $N_\mathrm{selected}$ is the number of galaxies passing a given set of cuts within the HSC wide catalog used, and $A$ is the sky area of the HSC field ($\sim$16 $\rm{{deg}^{2}}$, ignoring masked-out regions that likely reduce the effective area by a few percent). Finally, we define our net surface density yield to be
\begin{equation} \label{eq:density_yield}
\Sigma_\mathrm{yield}^{z\,= \, 1.1\, -\, 1.6}= \Sigma_\mathrm{target}\times f_\mathrm{z\,= \, 1.1\, -\, 1.6}^{700 \,<\,t\, <\,1400} 
\end{equation} .

We used a minimization method to optimize a set of free parameters defining our final selection cuts. The four parameters include the two color cut offsets for the color cuts obtained using the RF classifier, an $i-z$ cut (the next most important feature in the RF feature permutation), and a $g$-fiber limiting magnitude applied to the sample. As described in Section \ref{sec:forecasts}, the largest gains in enlarging the DESI ELG sample would be at high-z (where increasing past 1370 $\mathrm{deg^{-2}}$ would only result in marginal improvement to BAO measurements). Given this, we optimized our sample to maximize the redshift range success rate of high-z ELGs and obtain a net surface density yield close to 1370 $\mathrm{deg^{-2}}$.  We employed the \texttt{scipy.optimize.minimize} routine \citep{Scipy} for this purpose. Specifically, we minimized a loss function $L$ defined by: \begin{equation} \label{eq:loss}
L = -f_\mathrm{z\,= \, 1.1\, -\, 1.6}^{700 \,<\,t\, <\,1400}\times100\, +\, w\times(\Sigma_\mathrm{yield}^{z\,= \, 1.1\, -\, 1.6}- 1370 \, \mathrm{deg^{-2}})^{2},
\end{equation} 
where $w$ is a weight value within the range from $0.003$ to $0.0012$. After providing initial guesses for the optimal values of the four free parameters, we were able to extract values that maximized redshift range success rate with a net surface density yield close to our goal of 1370 $\mathrm{deg^{-2}}$. The final cuts obtained for our optimized sample are:

\begin{compactitem} \itemsep \listitemsep
\item $g_\mathrm{fiber} < 24.33$,
\item $i-y - 0.16 > r-i$,
\item $i-y > 0.43$, and 
\item $i-z > 0.45$.
\label{list:opt_cuts_truth}
\end{compactitem} The $i$ and $z$ bands proved instrumental in selecting high-z ELGs because they spanned the wavelengths where the 4000 \AA\, break moves between at $z\sim1.1$.

This work started before the conception of the spec-truth sample, so our initial efforts involved developing target selection cuts for a similar sample using LSST-like $ugrizy$ photometry and many-band photo-z's from COSMOS2020. We used these cuts to obtain DESI spectroscopy to test these selection criteria, which we refer to as the DESI-II Pilot Sample (see Appendix \ref{appendix:pilot_survey}). However, the spec-truth sample superseded the DESI-II Pilot Sample due to its larger size and broader color cuts, so we used it for all the results within the main body of this paper.
\subsection{Comparison to DESI}
\label{subsec:stats comparison}

\begin{figure*}
    \centering
    \includegraphics[width=\textwidth]{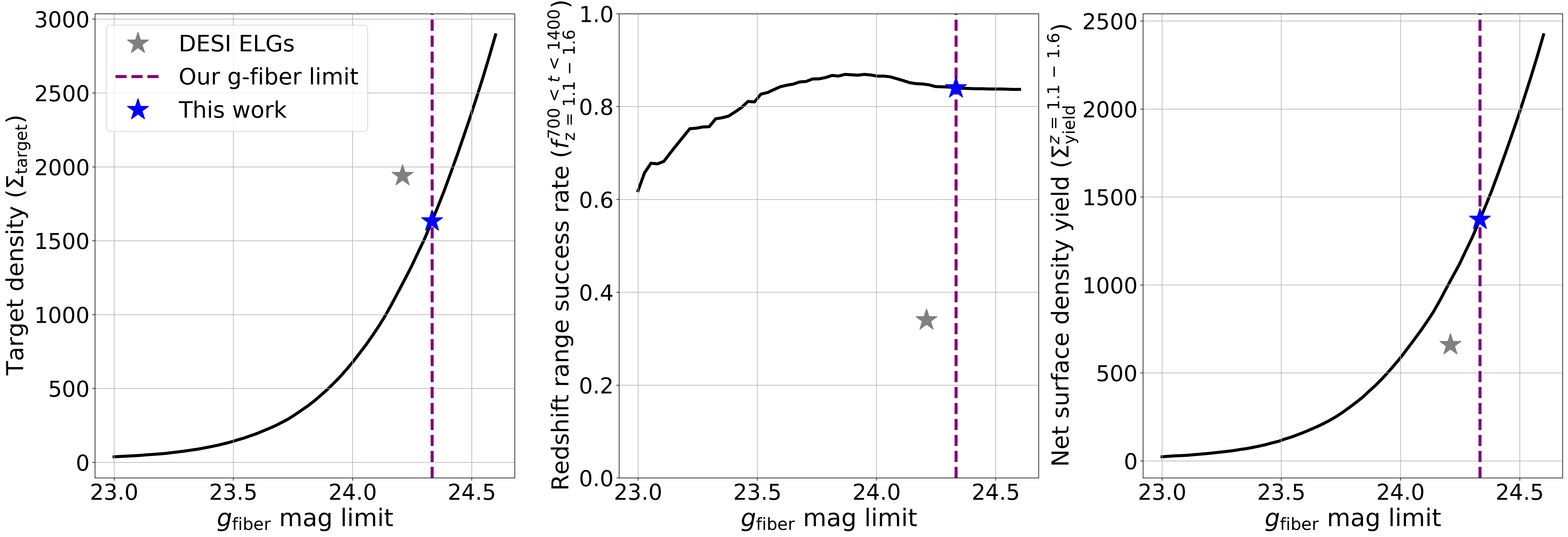}
    \caption{Target density ($\Sigma_\mathrm{target}$), redshift range success rate ($f_\mathrm{z\,= \, 1.1\, -\, 1.6}^{700 \,<\,t\, <\,1400}$), and the net surface density yield ($\Sigma_\mathrm{yield}^{z\,= \, 1.1\, -\, 1.6}$) for different $g$-fiber limiting magnitudes (black solid lines). The $g$-fiber limiting magnitude optimized to yield a net surface density of 1372 redshifts per square degree is indicated by the dashed purple lines in each panel. The DESI ELG sample is shown by the gray stars while the results of our sample are shown by the blue stars. Despite both samples having similar target densities and $g$-fiber limiting magnitudes, the HSC-based selection presented here is much more efficient at yielding redshifts in the desired range, as can be seen by the much higher values of redshift range success rate (middle panel) and net surface density yield (right panel) obtained for it.}
    \label{fig:3panel_sumstat}
\end{figure*}

The ELG sample presented here is optimized to cover the redshift range $1.1 < z < 1.6$ in order to allow direct comparisons to the statistics for the DESI ELG sample in this $z$ range that are presented in Table 4 of \citet{Raichoor_2023}. In Figure \ref{fig:3panel_sumstat}, we plot the target surface density ($\Sigma_\mathrm{target}$), the redshift range success rate ($f_\mathrm{z\,= \, 1.1\, -\, 1.6}^{700 \,<\,t\, <\,1400}$), and the net surface density yield ($\Sigma_\mathrm{yield}^{z\,= \, 1.1\, -\, 1.6}$) resulting from our color cuts as a function of the $g$-fiber limiting magnitude. The $g$-fiber magnitude limit we adopt as standard is indicated by the vertical purple dashed line; our sample can be compared directly to the statistics for DESI ELGs from \citet{Raichoor_2023} that are indicated by red stars. Despite having a similar target density and only going slightly fainter than the DESI ELG sample, our color cuts provide a much denser sample of objects with redshifts in the desired range. We compare our results to DESI ELGs quantitatively in Table \ref{stat_table}, using the metrics defined in Subsection \ref{subsec:colorcut_offset}. We find that the largest improvements are obtained for redshift range success rate and net surface density yield; our sample outperforms the DESI ELG sample by factors of $\sim2.4$ and $\sim2.1$ in these metrics, respectively.

\begin{table*}[!t]
\centering
\caption{Comparison of metrics for our ELG selection to those for DESI ELGs. By selecting ELGs with deeper photometry in redder bands, we can achieve more than a factor of two higher fraction of targets with reliable redshift measurements in the desired redshift range ($f_\mathrm{z\,= \, 1.1\, -\, 1.6}^{700 \,<\,t\, <\,1400}$). Because of this, we are able to more than double the net surface density yield ($\Sigma_\mathrm{yield}^{z\,= \, 1.1\, -\, 1.6}$) while targeting fewer total targets per square degree than the DESI ELG sample. 
 \label{stat_table}}
\label{tab:table_stats}
\begin{tabular}{|c|c|c|c|c|}
\hline
 & $f_\mathrm{reliable}$ 
 & $f_{z = 1.1 - 1.6}^{700 < t < 1400}$ 
 & $\Sigma_\mathrm{target}\,(\mathrm{deg}^{-2})$
 & $\Sigma_\mathrm{yield}^{z\,= \, 1.1\, -\, 1.6}\,(\mathrm{deg}^{-2})$ \\
\hline
DESI ELGs & 69\% & 34\% & 1940 & 660 \\
\hline
This work & 89\% & 84\% & 1632 & 1372 \\
\hline
\end{tabular}
\end{table*}

\begin{figure*}
    \centering
    \includegraphics[width=\textwidth]{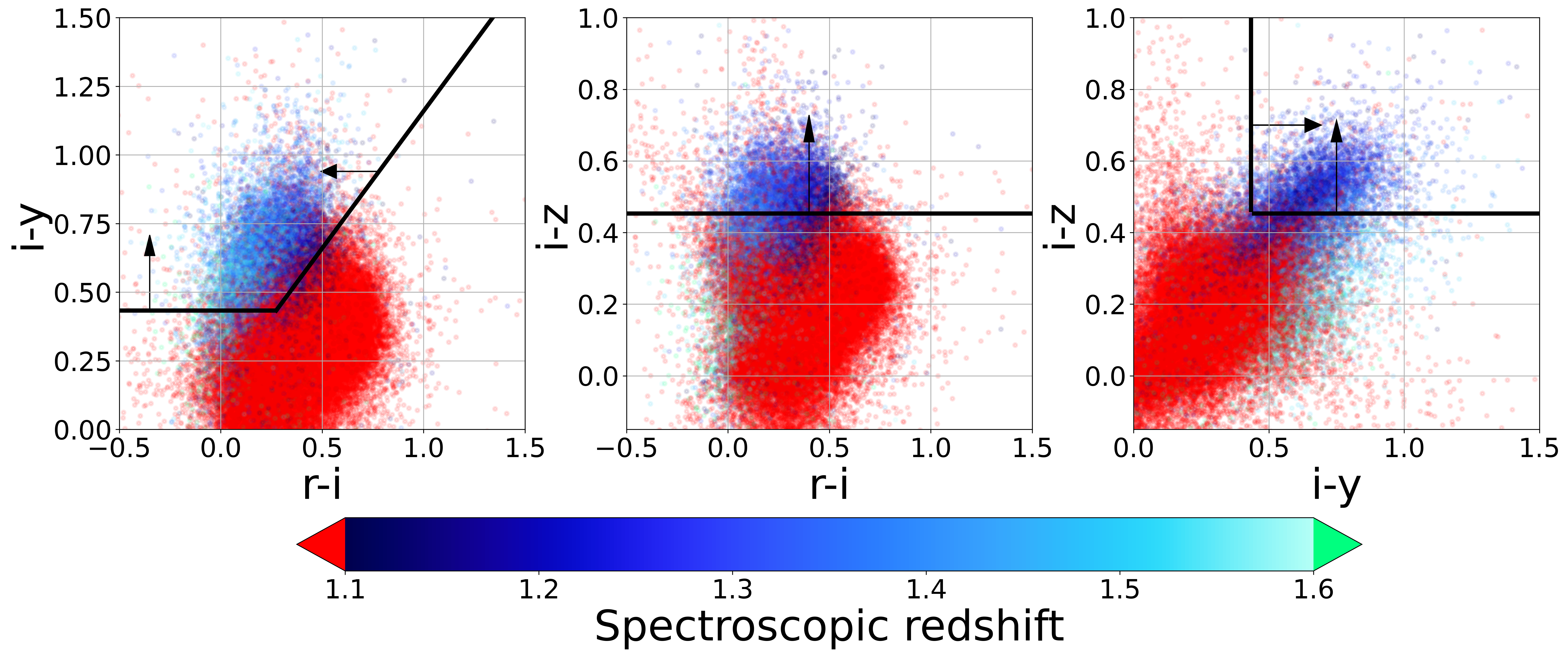}
    \caption{Color-color diagrams of ELGs from the spec-truth sample with reliable $z_\mathrm{spec}$ measurements, where points are color coded according to their spectroscopic redshifts.  Solid black lines show our final selection cuts, with black arrows indicating on which side of the dividing lines our final sample lies. These cuts are effective at throwing out objects at low redshift (the $z < 1.1$ objects indicated by red points) while maintaining a high completeness for selecting $1.1 < z < 1.6$ galaxies.}
    \label{fig:color_color_truth}
\end{figure*}

Figure \ref{fig:color_color_truth} illustrates the effectiveness of our optimized $r-i/i-y/i-z$ color cuts for selecting ELGs within the redshift range $1.1 < z < 1.6$. Our cuts reject the majority of the low redshift contamination at $z < 1.1$ (red), while still retaining the galaxies we wish to include, i.e., those which are at $1.1 < z < 1.6$ and yield secure redshifts in $\sim 1000$ second exposure times (blue points). 

\begin{figure*}
    \centering
    \includegraphics[width=\textwidth]{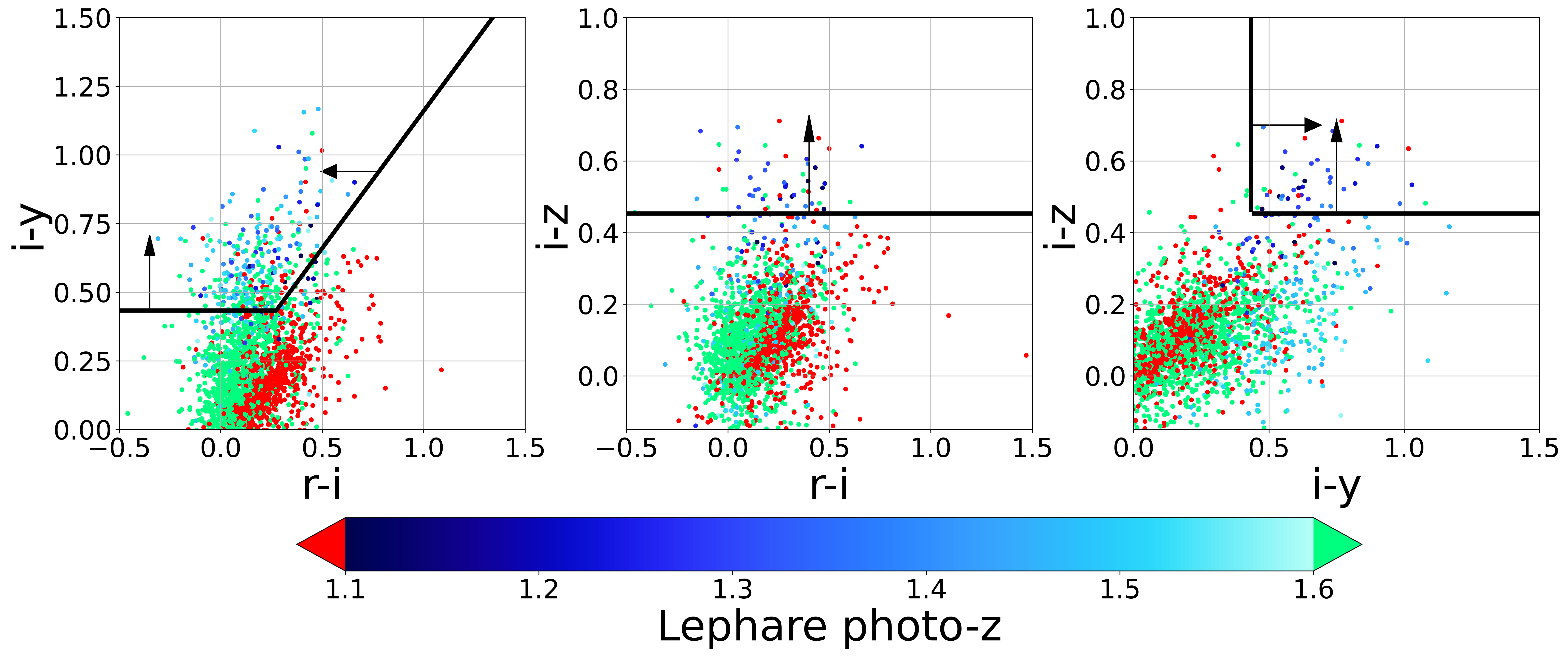}
    \caption{Color-color plot diagrams for galaxies from the spec-truth sample that did not meet the reliable redshift cuts illustrated in Figure \ref{fig:good_z_criteria}, with points color coded based upon their COSMOS2020 LePHARE many-band photo-z.  Solid black lines and arrows show our optimized color cuts. Our results indicate that the color cuts are not only extremely effective at throwing out redshift failures that intrinsically lie below $z < 1.1$ (red) but also are effective in rejecting objects at $z > 1.6$, where the DESI spectrograph would be ineffective in measuring redshifts (green).}
    \label{fig:color_color_fail}
\end{figure*}

By using COSMOS2020 LePHARE photo-z's, we can also explore how galaxies that did not pass our reliable redshift criteria populate this color space. Color-color plots for this sample color-coded according to their LePHARE photo-z's are shown in Figure \ref{fig:color_color_fail}. Although these photo-z's are not as reliable as spectroscopic redshift measurements, they still can provide useful information on the nature of the redshift failures. As can be seen in this plot, in addition to excluding low-z objects that are spectroscopically confirmed to be at $z < 1.1$ , our optimized cuts also are very effective at rejecting objects which failed to yield secure redshift measurements but are in fact estimated to lie either at $z < 1.1$ or at $z > 1.6$ (the regime where [O II] falls outside of DESI's spectral window so that redshift failures would be expected).

\begin{figure}
    \centering
    \includegraphics[width=\columnwidth]{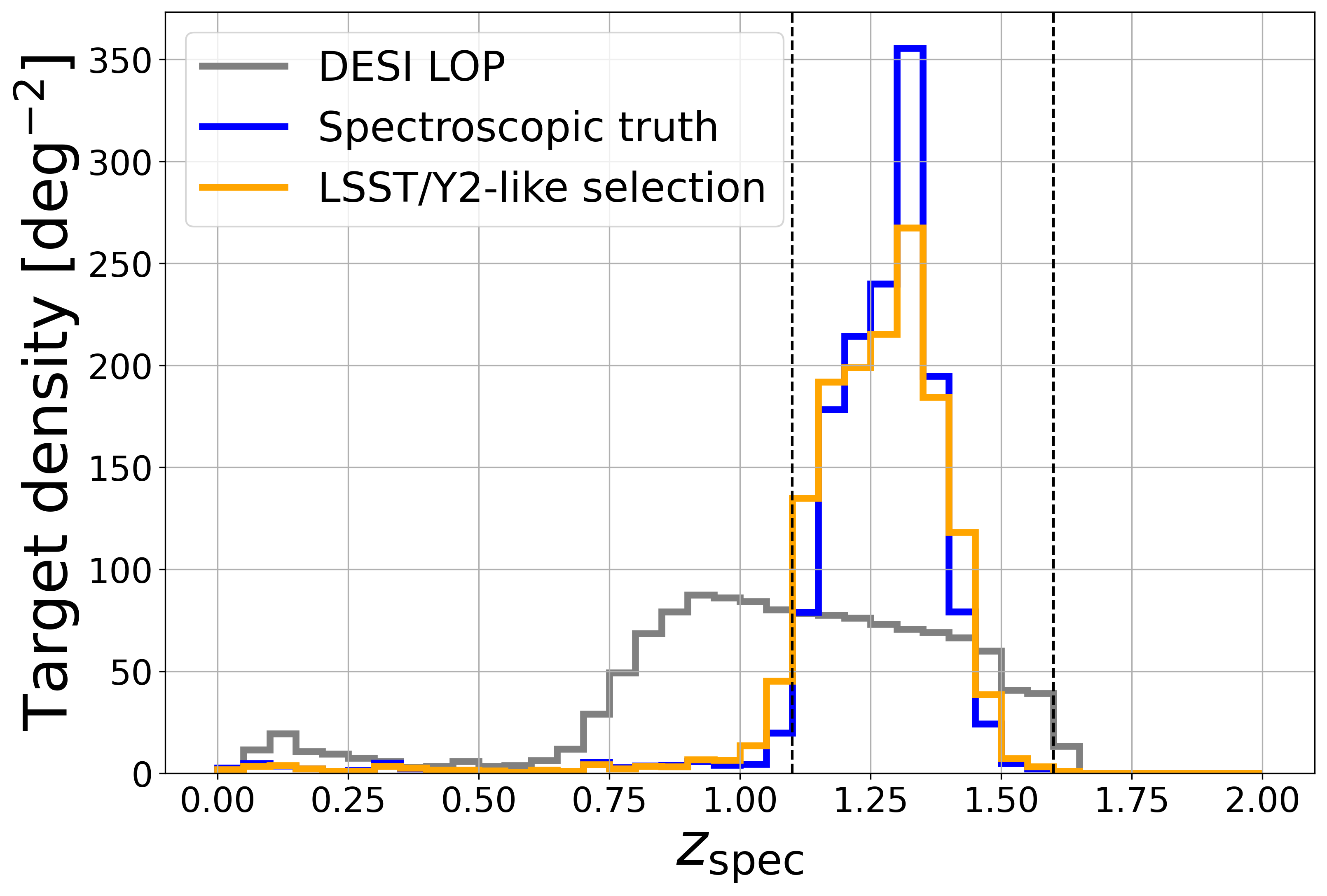}
        \caption{Comparison of the target density of objects as a function of redshift for the DESI ELG sample (gray) compared to the ELGs selected by the optimized cuts presented in Subsection \ref{subsec:colorcut_offset} (blue) and a similar selection using artificially degraded photometry (orange). Vertical lines indicate the redshift range $1.1 < z < 1.6$ which our sample is optimized to target. Our cuts do a much better job at excluding low redshift ($z < 1.1$) objects and consequently produce a much denser sample with galaxies in the desired redshift range for both samples. Rejecting low-z contamination around $z \sim0.35$ came at the cost of some high-z ELGs near $z \sim1.5$ (due to the two populations having similar colors).}
    \label{fig:elgs_compare_truth}
\end{figure}

In Figure \ref{fig:elgs_compare_truth}, we compare the distribution of spectroscopic measurements for our optimized sample and a similar sample using slightly shallower LSST/Y2-like photometry (described in Appendix \ref{appendix:shallower_selection}) directly to DESI ELGs. Both optimized samples suffer from less low-z ($z < 1.1$) contamination while also having a much larger net surface density yield of ELGs within the desired redshift range.

We have applied the same methods to optimize selections for a slightly broader redshift range, $1.05 < z < 1.65$, and obtained qualitatively similar results; i.e., our methods can be optimized for somewhat different redshift ranges if desired. Most notably, when optimizing for this broader span we obtain a slightly higher redshift range success rate ($f_\mathrm{z\,= \, 1.05\, -\, 1.65}^{700 \,<\,t\, <\,1400}$) of 85\% as compared to 84\% for the $1.1 < z < 1.6$ sample. 

Utilizing the deeper imaging and additional bands from COSMOS2020, we designed and optimized a high-z ELG sample that is denser and more efficient than current DESI ELGs. In particular, our sample produces a significantly higher redshift range success rate and net surface density yield by factors of $\sim2.4$ and $\sim2.1$ respectively. Of the spec-truth galaxies that passed our color and g-fiber magnitude cuts, only around 17.4\% were also LOP targets. Combining the current DESI ELG sample with ours would be instrumental in reducing BAO errors at $1.1 < z < 1.6$.

\section{Summary and Prospects}
\label{sec:Summary / Future  Tests}

In this work, we have demonstrated that current and planned deep multi-band imaging data sets can enable the design of efficient high-z ELG selections for future spectroscopic surveys such as a future DESI-II survey. As described in Section \ref{sec:forecasts}, by enlarging the current DESI ELG sample close to a factor of 3$\times$, we can reduce uncertainties of BAO distance parameters ($\alpha_\perp$ and $\alpha_\parallel$) for high-z ELGs by a factor of 2$\times$.

With this goal in mind, we explored different selection criteria for ELGs at $1.1 < z < 1.6$ using $grizy$ HSC wide photometry, a RF trained on COSMOS2020 $ugrizy$ photometry and LePHARE photo-z's, and results of a dedicated spectroscopic campaign with DESI. Using a simple minimization approach, we obtained a sample optimized for selecting high redshift ELGs that incorporates cuts in $r-i$, $i-z$, and $i-y$ colors as well as a limit in $g$-band fiber magnitude.  Compared to DESI,
the sample has a similar target density ($\Sigma_\mathrm{target}$) and $g$-band fiber magnitude limit, yet results in a much denser and efficient ELG selection.  The largest improvements are made on redshift range success rate ($f_\mathrm{z\,= \, 1.1\, -\, 1.6}^{700 \,<\,t\, <\,1400}$) and net surface density yield ($\Sigma_\mathrm{yield}^{z\,= \, 1.1\, -\, 1.6}$), which our sample outperforms by over factors of $\sim2.4$ and $\sim2.1$, respectively. 

Combining the current ELG sample with ours would help reduce BAO errors at $1.1 < z < 1.6$ by around a factor of $\sim2$. We have also used the DESI spec-truth sample of ELG-like objects to test the performance of photo-z algorithms at predicting their redshifts, finding significant scatter and outliers compared to their spec-z (especially when comparing HSC photo-z's), reinforcing the need for combining photometric information with spectra when designing high-z ELG samples (see Appendix \ref{appendix:photoz_tests}).  

Looking towards the future, it would be interesting to test the feasibility of obtaining an ELG sample like the one described in this work during gray or bright time, rather than the comparatively dark time that was used for the spec-truth observations. Moonlight mostly affects the night sky continuum spectrum, particularly at the blue end where scattering is greatest.  As a result, it should have comparatively little effect on redshift success for our sample, as at $z>1.1$ the [OII] line will be at $>7800$ \AA \,where night-sky emission lines are the dominant background. Obtaining such a sample during gray or bright time could increase the efficiency of a DESI-II program, as its main cosmological samples will focus on $z>2$ galaxies, for which the primary features of interest are at the blue end of the spectrum and dark time is necessary.

Another application of the sample selection described here (or a similar selection) would be to produce a purely photometric sample with a well-defined selection and redshift distribution that could be used for CMB cross-correlation analyses.  Such samples would build on the photometric LRG selections presented by \citet{Zhou_2023_cc}, which enable such analyses at lower redshifts. For measurements of lensing cross-correlations, precision redshift measurements for individual objects are not necessary so long as $z$ distributions are tight and well-understood. Defining such photometric samples would enable constraints on the growth of structure at higher redshifts than LRGs probe, extending the work of \citet{Sailer_2025} past $ z \sim 1$.

\section{Acknowledgements}
The efforts of Y. Salcedo Hernandez and J. A. Newman were supported by grant DE-SC0007914 from the U.S. Department of Energy Office of Science, Office of High Energy Physics.

This material is based upon work supported by the U.S. Department of Energy (DOE), Office of Science, Office of High-Energy Physics, under Contract No. DE–AC02–05CH11231, and by the National Energy Research Scientific Computing Center, a DOE Office of Science User Facility under the same contract. Additional support for DESI was provided by the U.S. National Science Foundation (NSF), Division of Astronomical Sciences under Contract No. AST-0950945 to the NSF’s National Optical-Infrared Astronomy Research Laboratory; the Science and Technology Facilities Council of the United Kingdom; the Gordon and Betty Moore Foundation; the Heising-Simons Foundation; the French Alternative Energies and Atomic Energy Commission (CEA); the National Council of Humanities, Science and Technology of Mexico (CONAHCYT); the Ministry of Science, Innovation and Universities of Spain (MICIU/AEI/10.13039/501100011033), and by the DESI Member Institutions: \url{https://www.desi.lbl.gov/collaborating-institutions}. Any opinions, findings, and conclusions or recommendations expressed in this material are those of the author(s) and do not necessarily reflect the views of the U. S. National Science Foundation, the U. S. Department of Energy, or any of the listed funding agencies.

The authors are honored to be permitted to conduct scientific research on I'oligam Du'ag (Kitt Peak), a mountain with particular significance to the Tohono O’odham Nation.

Y. Salcedo Hernandez thanks the LSST-DA Data Science Fellowship Program, which is funded by LSST-DA, the Brinson Foundation, the WoodNext Foundation, and the Research Corporation for Science Advancement Foundation; his participation in the program has benefited this work.

\software{Astropy \citep{The_Astropy_Collaboration_2022}, Colorcet \citep{kovesi2015goodcolourmapsdesign},
Matplotlib \citep{Hunter:2007}, NumPy \citep{harris2020array}, SciPy \citep{2020SciPy-NMeth}, SciKit-Learn, \citep{scikit-learn}, Pandas, \citep{reback2020pandas}
          }


\section{Data Availability}
The data used to make the plots in this paper can be found at \url{https://zenodo.org/records/18702588}

\appendix
\label{appendix}

\section{DESI-II pilot sample}
\label{appendix:pilot_survey}

Prior to the availability of the DESI spec-truth sample, we used the DESI instrument to obtain spectroscopy of a sample of objects intended to investigate the feasibility of using deep wide-area multi-band imaging (e.g., from LSST) to efficiently select high redshift ELGs; we will refer to this as the ``pilot sample'' in the remainder of this Appendix. Analyses of this sample led to similar conclusions to the results based upon the spec-truth sample reported in the main body of this paper. Given its broader targeting color cuts and larger sample size (with over a 3$\times$ increase in the number of high-z ELGs), we have used the spec-truth sample for the main analysis of this paper; however, we briefly describe the pilot sample here to document its selection and its implications for high-z ELG selection. This sample should be included in future public DESI data releases.


\subsection{Pilot Sample Design}
\begin{figure*}[!h]
    \centering
    \includegraphics[width=\textwidth]{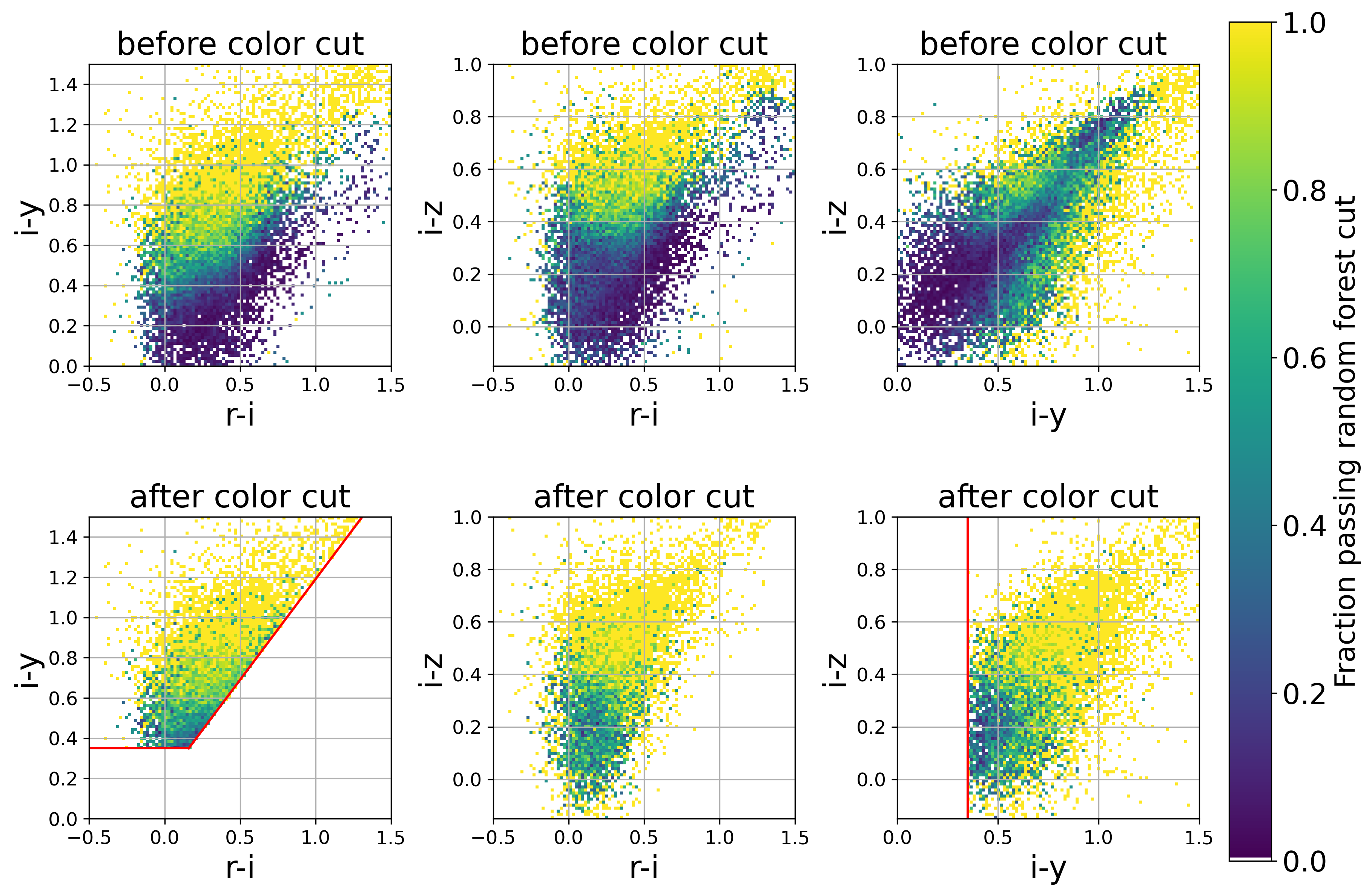}
    \caption{Color-color plots of $r-i$/$i-y$/$i-z$ color space, showing the fraction of objects that pass our RF probability $ \geq 0.025$ cut before (top row of panels) or after (bottom panels) the simple color cuts used for targeting the pilot sample of potential high-redshift ELGs are applied. This figure includes only HSC wide galaxies with COSMOS2020 cross-matches. Guided by the RF, we were able to design simple color cuts which are effective at selecting objects within the redshift range of interest ($1.1 < z < 1.6$).}
    \label{fig:colorcut_rf_pass_pilot}
\end{figure*}

Using HSC wide photometry, COSMOS2020 photometric redshifts, and the aid of a RF classifier to guide the choice of colors to use (following the procedures that are described in detail in Subsection \ref{subsec:color_cut_design}), we developed a set of simple $r-i$/$i-y$ color cuts that could be used to target galaxies likely to be high-redshift ELGs. The resulting cuts are illustrated in Figure \ref{fig:colorcut_rf_pass_pilot}. Galaxies in the color-color plots in each panel are color coded according to the fraction that pass an RF cut designed to select objects within $1.1 < z <  1.6$. This RF classifier, like the one in Section \ref{subsec:color_cut_design}, was trained on all magnitudes and color combinations from COSMOS2020.

The divisions used for targeting were originally designed to select galaxies within the redshift range $1.05 < z < 1.55$, but in order to facilitate direct comparisons to statistics for the DESI ELG sample, we used the spectroscopic redshifts from the pilot sample to develop optimized cuts for selecting objects in the $1.1 < z < 1.6$ redshift range. 

The cuts applied for the targeting of the pilot sample were shifted slightly from the RF optimized values to avoid regions which HSC photo-z's suggested were dominated by lower-redshift galaxies, resulting in a final set of selection cuts that were more exclusive in $r-i/i-y$ than the values optimized based on COSMOS2020 photometric redshifts and the classifier. The excluded portion of color space is included in our final color cuts which were optimized based upon the spec-truth sample, reflecting the impact of the catastrophic failures for HSC photo-z's of ELGs that are illustrated in Figure \ref{fig:phot/specz cosmos}. 

We utilized a cutoff of 0.025 in the RF probability to distinguish the locations of low-redshift from high-redshift objects in color space; this choice was tuned based on the ROC curve to provide a high completeness for selecting high-z objects while maintaining a low degree of contamination.  
The final targeting cuts used for the pilot survey spectroscopic sample were as follows:

\begin{compactitem} \itemsep \listitemsep
\item $g$ band magnitude limits: $(g_\mathrm{fiber} < 24.3)$ \textbf{AND} $(g < 24)$;
\item $r$ band magnitude limits: $(r_\mathrm{fiber} < 24.3)$ \textbf{AND} $(r < 24)$; and
\item color cut: $(i-y -0.19 > r-i)$  \textbf{AND} $(i-y > 0.35)$ 
\end{compactitem}.

All objects which fulfilled either of the $g$ or $r$ band magnitude limits, and also fulfilled the sample color cuts, were included in the target sample. In total, spectra of 22,089 objects following these selection cuts were obtained in the pilot sample. Of those objects we obtained spectra for, 20,646 had reliable redshift measurements.

\begin{figure}
    \centering
    \includegraphics[width=\columnwidth]{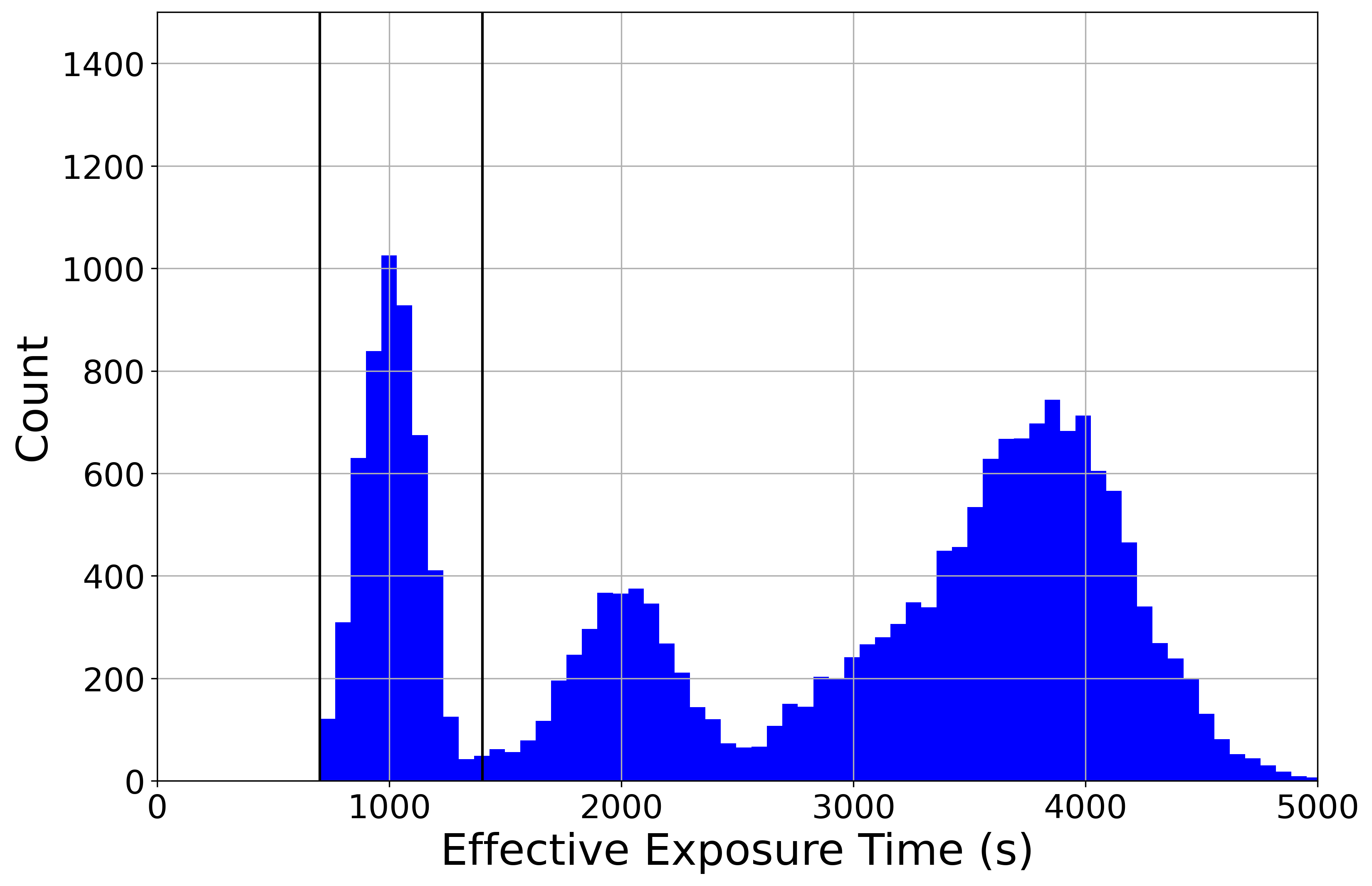}
    \caption{The exposure time distribution for the pilot sample. Only objects with effective exposure times in the range $700 < t < 1400$ seconds (indicated by the vertical black lines) were used for calculating redshift measurement success rates ($f_\mathrm{reliable}$), whereas objects with longer exposures were included when calculating the redshift range success rate ($f_\mathrm{z\,= \, 1.1\, -\, 1.6}^{700 \,<\,t\, <\,1400}$).}
    \label{fig:exposure_pilot}
\end{figure}

In Figure \ref{fig:exposure_pilot}, we show the distribution of effective exposure time (cf. Section \ref{sec:observations}) for the pilot sample galaxies. Only objects with exposure times in the range $700 < t_{\rm effective} < 1400$ seconds (between the solid black vertical lines on the figure) were used to calculate redshift measurement success rates in order to enable fair comparisons to the DESI ELG sample, for which typical effective exposure times are $\simeq$1000 seconds. 
Objects with longer exposures were, however, included when calculating redshift distributions and related statistics in order to maximize sample sizes. 

\subsection{Results}

\begin{figure}
    \centering
    \includegraphics[width=\textwidth]{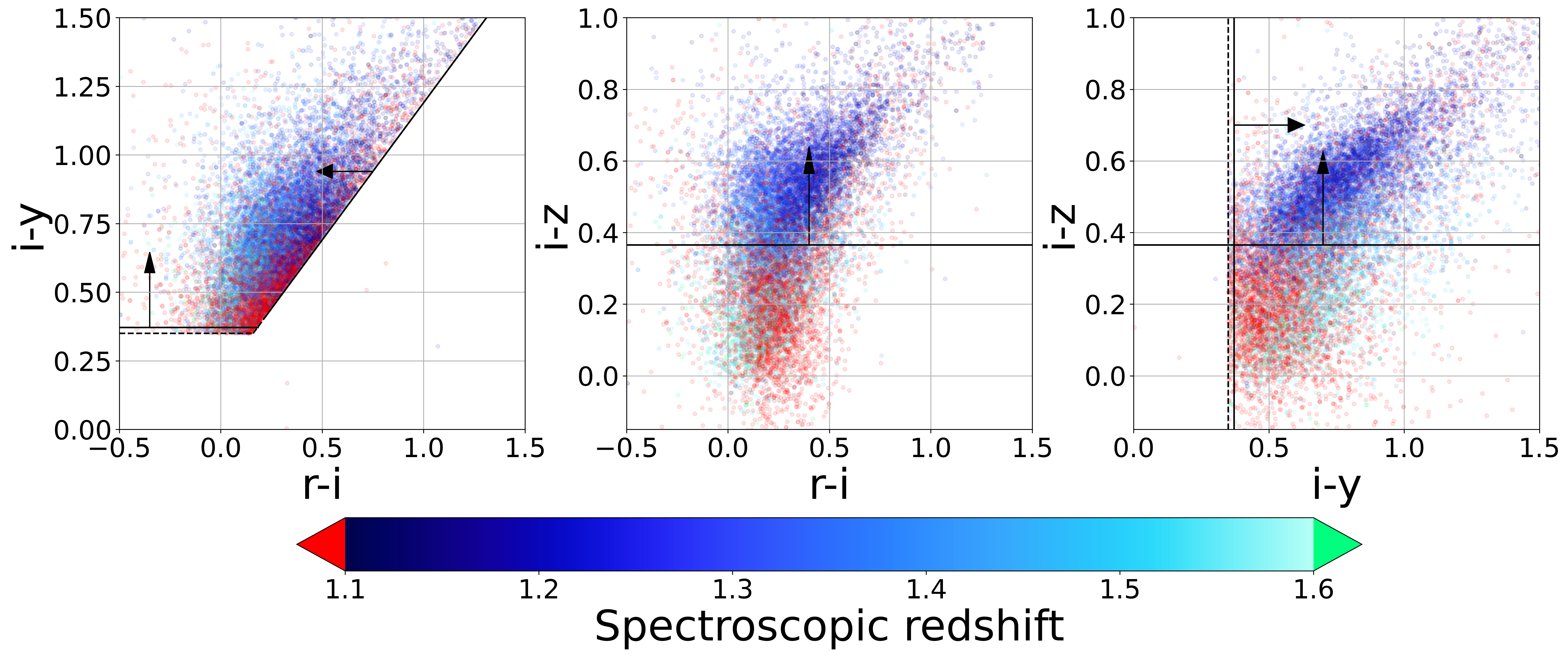}
    \caption{Color-color plots of galaxies with reliable redshifts from the high-z ELG sample in $r-i/i-y/i-z$ space, color-coded according to their spectroscopic redshift. Black dashed lines mark the color cuts used for targeting and solid lines the final cuts optimized based upon the spectroscopy. Adding an $i-z$ color cut was extremely effective at excluding $z < 1.1$ galaxies. The few objects beyond the limits of the dashed lines were due to inconsistencies between photometric catalogs.}
    \label{fig:color_color_pilot}
\end{figure}

Using the same optimization methods to fine-tune the final selection cuts as those described in Subsection \ref{subsec:colorcut_offset}, we were able to select a subset of the pilot survey sample which yielded values of $f_\mathrm{reliable}$, $f_\mathrm{z\,= \, 1.1\, -\, 1.6}^{700 \,<\,t\, <\,1400}$, and $\Sigma_\mathrm{yield}^{z\,= \, 1.1\, -\, 1.6}$ of 89\%, 77\%, and 1375 $\mathrm{deg^{-2}}$, respectively. The final cuts obtained for our pilot survey sample are:

\begin{compactitem} \itemsep \listitemsep
\item $g_\mathrm{fiber} < 24.22$,
\item $i-y - 0.19 > r-i$,
\item $i-y > 0.37$, and 
\item $i-z > 0.37$.
\label{list:opt_cuts_pilot}
\end{compactitem}
Comparing to the sample optimized based upon the spec-truth observations, the most noticeable difference is the lower redshift range success rate (77\%\ versus\ 84\%) attained here. This difference is likely due to our inability to explore the $r-i$/$i-y$ color space as freely due to the stricter color cuts used for targeting (see the first panel of the bottom row in Figure \ref{fig:colorcut_rf_pass_pilot}). The final color cuts optimized based upon this sample are shown in Figure \ref{fig:color_color_pilot}. These divisions are very similar to those obtained based upon the spec-truth sample; however, due to the restriction applied to the pilot sample targets based upon HSC photo-z predictions, it was not possible to test positive offsets to the $r-i/i-y$ diagonal line, limiting the degree of optimization possible. 

\begin{figure}
    \centering
    \includegraphics[width=\textwidth]{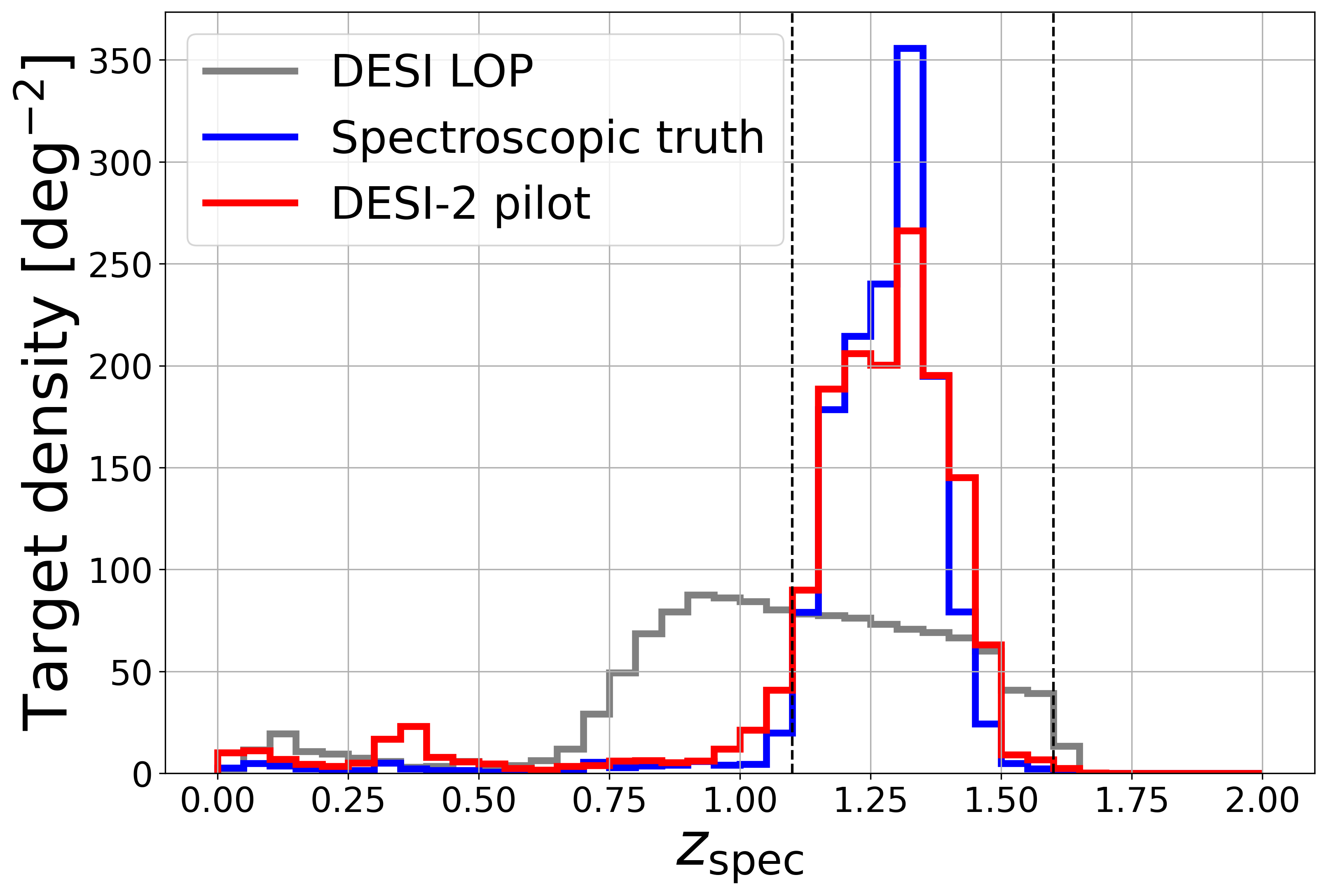}
    \caption{Redshift distributions (in galaxies per square degree per redshift bin) for DESI ELGs (grey), the high-z ELG sample optimized based upon the spec-truth observations (blue), and the high-z ELG sample optimized  based upon the pilot sample spectroscopy (red). Unlike the DESI ELGs, both the pilot and spec-truth-optimized samples are almost entirely within the redshift range of greatest interest, $1.1 < z < 1.6$. Both samples are also much more efficient at selecting high-redshift ELGs than DESI ELGs, thanks to being able to exploit deeper photometry in more bands than were used for DESI.}
    \label{fig:elgs_compare_pilot}
\end{figure}

In Figure \ref{fig:elgs_compare_pilot} we compare the redshift distribution of the ELG sample optimized based on the pilot survey observations to both the DESI ELGs and the ELG sample optimized based upon the spec-truth spectra. As with the spec-truth-optimized sample, the pilot survey observations demonstrate that deeper multiband imaging makes it possible to significantly reduce low-redshift contamination and acquire a denser sample over the desired  $1.1 < z < 1.6$ redshift range compared to DESI. Although it exhibits slightly more contamination at $z < 1.1$, the pilot sample optimization provides qualitatively similar results to the spec-truth selection. Combining the current DESI ELGs with samples designed with either of these sets of color cuts would allow for a future spectroscopic survey to provide significantly improved constraints on BAO cosmology at redshifts $1.1 < z < 1.6$.


\section{Impact of shallower photometry}
\label{appendix:shallower_selection}

\subsection{Methods}

The HSC wide imaging is slightly deeper in some bands than the photometry that will be available early in the LSST survey.  In order to test how being limited to shallower photometry might impact the selection of high-redshift ELGs, we have designed a set of selection cuts based upon artificially degraded $grizy$ photometry from HSC wide.

First, we take the $5\sigma$ magnitude limits that LSST will achieve after the full ten-year (Y10) dataset and degrade them to the expected values for a two-year (Y2) dataset, assuming that signal-to-noise scales as the square root of total exposure time as expected in the background-limited regime: 
\begin{equation} 
\label{eq:year_two_depth}
m_\mathrm{lim,LSST \, Y2} = m_\mathrm{lim,LSST \, Y10} + 1.25\log(2/10), 
\end{equation} 
where $m_\mathrm{lim, LSST \, Y2}$ and $m_\mathrm{lim, LSST \, Y10}$ are the Y2 and Y10 $5\sigma$ magnitude limits, respectively \citep{Bianco_2021}. Using these predicted magnitude limits, we can calculate the expected errors for a particular object according to the equation:
\begin{equation} 
\label{eq:sigma_m}
\sigma_\mathrm{m} = \frac{1}{2 \ln{10}} \times10^\frac{m_\mathrm{HSC\,wide }\,-\, m_\mathrm{lim, LSST\,Y2}}{2.5}, 
\end{equation} 
where $m_\mathrm{HSC\,wide}$ is the HSC wide magnitude for that object in a particular band and $m_\mathrm{lim, LSST\,Y2}$ is the LSST Y2 5$\sigma$ depth in the corresponding band. One can derive this equation via propagation of error from errors in flux into errors in magnitude, combined with the requirement that the fractional flux error must be one-fifth when $m_\mathrm{HSC\,wide } =  m_\mathrm{lim, LSST\,Y2}$. We compare the 5$\sigma$ depths for all bands of HSC DR3 to what is expected for LSST Y2 in Table \ref{depth_comparisons}. Dust can have a small effect on the effective depths, especially for the bluest bands, but we ignore that here.

\begin{table}[!h]
\label{fig:depth_comparisons}
\begin{center} 
    \begin{tabular}{|c|c|c|c|c|c|}
\hline
5$\sigma$ depths & $g$ & $r$ & $i$& $z$ & $y$ \\
\hline
HSC DR3 & 26.5 & 26.5 & 26.2 & 25.2 & 24.4 \\
\hline
LSST Y2 & 26.5 & 26.6 & 25.9 & 25.2 & 24.0 \\
\hline
\end{tabular}
\end{center}

\caption{5$\sigma$ PSF depths of HSC DR3 and LSST after two years of observations. The biggest difference between the two are for the $y$ band, where HSC is deeper by 0.4 mag. 
 \label{depth_comparisons}}
\end{table}

We then generate random values from Gaussians with a mean of zero and a standard deviation of the appropriate $\sigma_\mathrm{m}$ value to each object's magnitudes, and add them to the original HSC magnitudes to obtain a set of noisier $grizy$ measurements for all objects and bands. This gives us a set of magnitude values which scatter about the original HSC photometry at a level corresponding to Equation \ref{eq:sigma_m}.
It is important to note that this procedure results in over-degraded photometry, since the HSC photometry we apply expected LSST errors to has non-zero observational uncertainties, making our tests conservative.

\subsection{Results}
\begin{figure}[!h]
    \includegraphics[width=\textwidth]{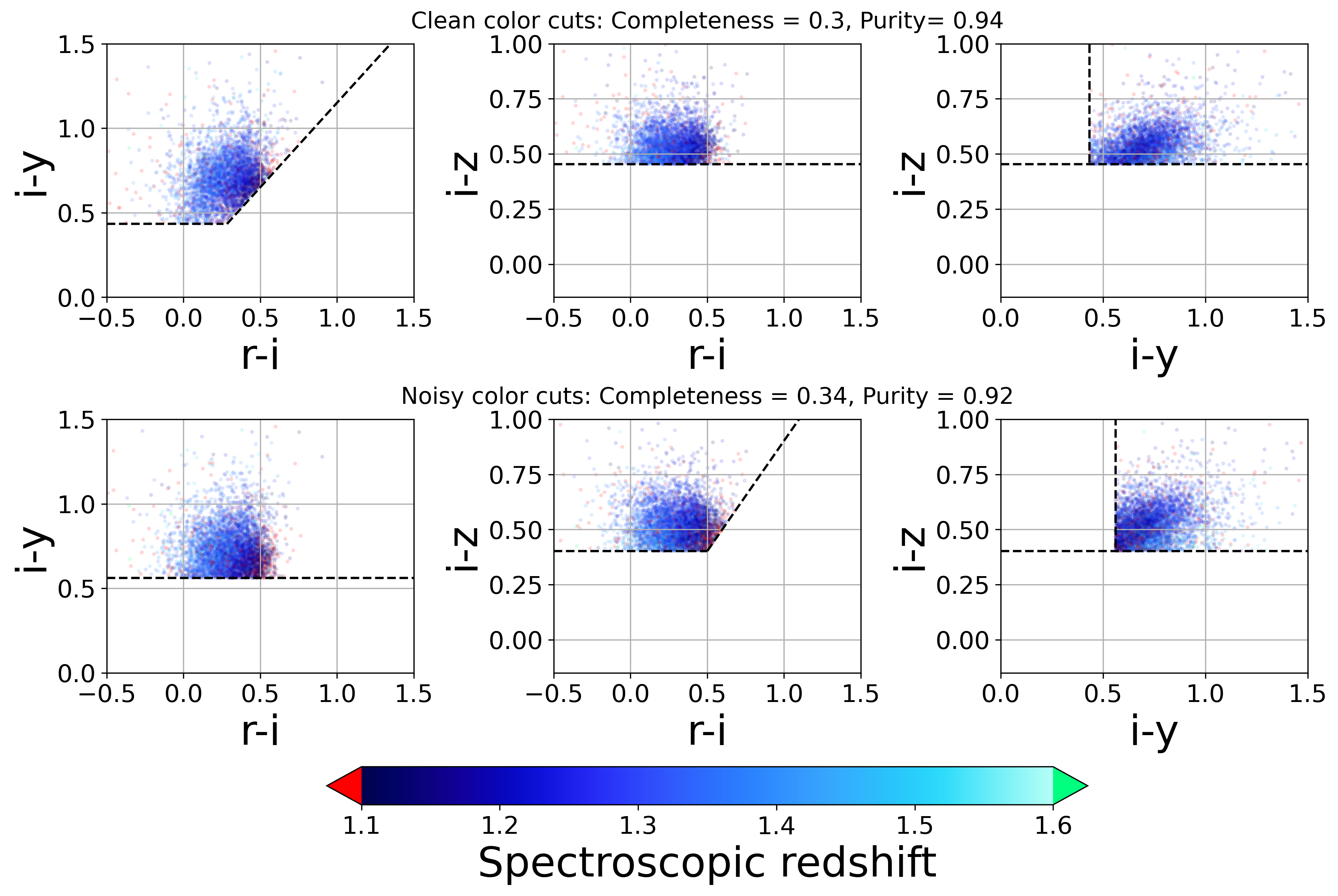}
    \caption{Color-color diagram of spec-truth galaxies color-coded by their $z_\mathrm{spec}$ with the positions of objects in color-color space plotted using either the original HSC wide photometry (top panel)  or their positions after photometry is degraded based on LSST Y2 depths (bottom) after applying optimized color cuts. Despite a small decrease in purity, we are still able to obtain a very clean sample of galaxies that are predominantly in the redshift range $1.1 < z < 1.6$.}
    \label{fig:noise_colorcut}
\end{figure}

    After obtaining this degraded photometry, we applied the same methods described in Section \ref{sec:optimization} to optimize selections for a sample of ELGs within the redshift range $1.1 < z < 1.6$ based upon the spec-truth redshifts. We compare these cuts optimized for noisier photometry to those we obtained before in Figure \ref{fig:noise_colorcut}. After adding noise to the photometry, a diagonal $i-y/r-i$ cut became less ideal and we went with a diagonal cut in $i-z/r-i$ instead. It is clear that it is still possible to obtain an efficient and dense selection of ELGs over the redshift range $1.1 < z < 1.6$ when limited to this noisier photometry. When optimizing based on the noisy photometry, we obtain values of  redshift measurement success rate ($f_\mathrm{reliable}$) of 89\% (as compared to 89\% for the sample described in Section \ref{subsec:colorcut_offset}), redshift range success rate ($f_\mathrm{z\,= \, 1.1\, -\, 1.6}^{700 \,<\,t\, <\,1400}$) of 83\% (versus 84\%), a target density ($\Sigma_\mathrm{target}$) of 1656 $\mathrm{deg^{-2}}$ (compared to 1632 $\mathrm{deg^{-2}}$) and net surface density yield ($\Sigma_\mathrm{yield}^{z\,= \, 1.1\, -\, 1.6}$) of 1366 $\mathrm{deg^{-2}}$ (versus 1372 $\mathrm{deg^{-2}}$) respectively. Our results are qualitatively similar to those achieved with the non-degraded photometry while still greatly outperforming the DESI ELG sample, suggesting that LSST Y2 photometry would be sufficient to select ELG targets for a DESI-II program. The final cuts obtained for this noisier optimization are:

\begin{compactitem} \itemsep \listitemsep
\item $g_\mathrm{fiber} < 24.27$,
\item $i-z + 0.10 > r-i$,
\item $i-y > 0.56$, and 
\item $i-z > 0.40$.
\label{list:opt_cuts_noisy}
\end{compactitem}

\section{Testing photometric redshift performance for high-z ELGs}
\label{appendix:photoz_tests}

\begin{figure}[!h]
    \includegraphics[width=\columnwidth]{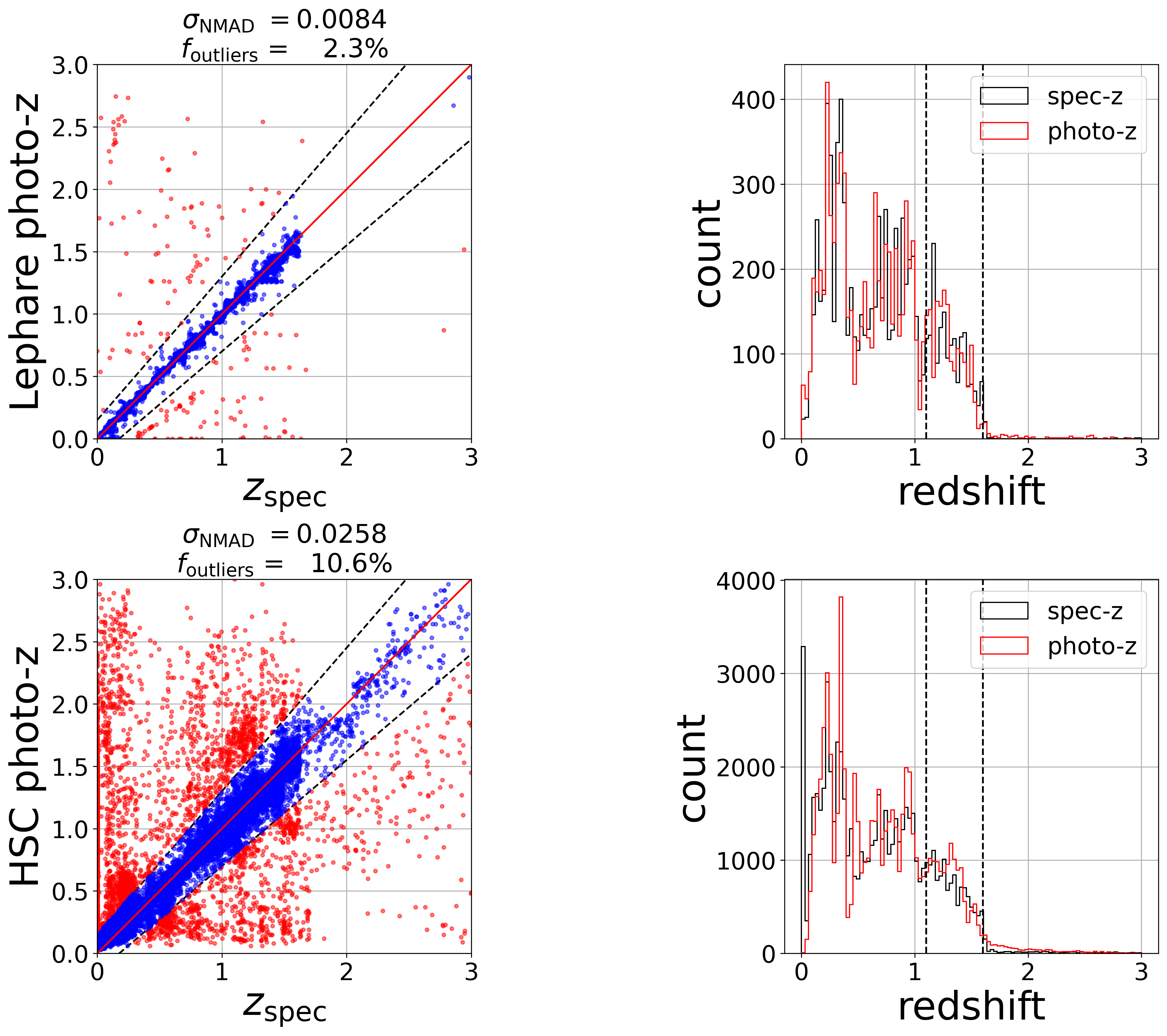}
    \caption{Comparison of COSMOS2020 (top row) or HSC (bottom row) photo-z's to the secure spectroscopic redshift measurements ($z_\mathrm{spec}$) for the spec-truth sample. Here, $f_\mathrm{outliers}$ is defined as the fraction with redshift difference $\frac{\Delta z}{1 + z} > 0.15$. The limited information available in broadband $grizy$ and the rarity of faint, low-redshift galaxies in training sets likely cause the much larger catastrophic outlier rates seen for the HSC photo-z's.}
    \label{fig:phot/specz cosmos}
\end{figure}

In order to test the accuracy and performance of the COSMOS2020 photo-z's used to design our initial ELG selections, we can compare them to the spectroscopic redshifts of the same objects. We have also tested the HSC DNNz $grizy$ photo-z's which are available for all of our objects \citep{Tanaka_2017, nishizawa2020photometricredshiftshypersuprimecam}. In Figure \ref{fig:phot/specz cosmos}, we show comparisons of the $z_\mathrm{spec}$ measurements for spec-truth objects with secure redshifts to either COSMOS2020 or HSC photo-z's, using both scatter plots and redshift histograms to compare.

Although it is only available for a fraction of the spec-truth sample due to the smaller sky area of the COSMOS field, the LePHARE photo-z's perform very well for this sample, having $\sigma_\mathrm{NMAD}$ and $f_\mathrm{outliers}$ of 0.0084 and 2.3\% respectively.  However, a handful of objects that are in fact at low redshift ($z\sim0.25$) are instead placed at much higher ($z\sim 2.5$). In contrast, the HSC $grizy$ photo-z's exhibit a much higher catastrophic outlier rate ($f_\mathrm{outliers} = 10.6$\%), particularly for objects that are in fact at $z\sim0.1$. Similar results are reported in Appendix B of \citet{Raichoor_2023}. The larger scatter and outlier rate is presumably driven by the limited information about redshift available from $grizy$ broadband photometry, though lack of faint, low-redshift galaxies in photo-z training sets may also play a role.  The systematic failures of photo-z algorithms in relevant regimes highlights the importance of obtaining spectra when designing final selection criteria for high redshift ELGs, as we have done in this work.

\bibliographystyle{aasjournalv7}
\bibliography{references}{}

\end{document}